\documentclass[twocolumn,showpacs, amsmath,amssymb,superscriptaddress,nofootinbib]{revtex4-1}
\usepackage{graphicx}
\usepackage{dcolumn}
\usepackage{bm}
\usepackage[chatter]{rotating}
\def\bea{\begin{eqnarray}}
\def\eea{\end{eqnarray}}

\begin{document}

\title{Measurement of D$^0$-meson + hadron two-dimensional angular correlations in Au+Au collisions at $\sqrt{s_{\rm NN}} = $ 200 GeV}

\affiliation{Abilene Christian University, Abilene, Texas   79699}
\affiliation{AGH University of Science and Technology, FPACS, Cracow 30-059, Poland}
\affiliation{Alikhanov Institute for Theoretical and Experimental Physics NRC "Kurchatov Institute", Moscow 117218, Russia}
\affiliation{Argonne National Laboratory, Argonne, Illinois 60439}
\affiliation{American University of Cairo, New Cairo 11835, New Cairo, Egypt}
\affiliation{Brookhaven National Laboratory, Upton, New York 11973}
\affiliation{University of California, Berkeley, California 94720}
\affiliation{University of California, Davis, California 95616}
\affiliation{University of California, Los Angeles, California 90095}
\affiliation{University of California, Riverside, California 92521}
\affiliation{Central China Normal University, Wuhan, Hubei 430079 }
\affiliation{University of Illinois at Chicago, Chicago, Illinois 60607}
\affiliation{Creighton University, Omaha, Nebraska 68178}
\affiliation{Czech Technical University in Prague, FNSPE, Prague 115 19, Czech Republic}
\affiliation{Technische Universit\"at Darmstadt, Darmstadt 64289, Germany}
\affiliation{ELTE E\"otv\"os Lor\'and University, Budapest, Hungary H-1117}
\affiliation{Frankfurt Institute for Advanced Studies FIAS, Frankfurt 60438, Germany}
\affiliation{Fudan University, Shanghai, 200433 }
\affiliation{University of Heidelberg, Heidelberg 69120, Germany }
\affiliation{University of Houston, Houston, Texas 77204}
\affiliation{Huzhou University, Huzhou, Zhejiang  313000}
\affiliation{Indian Institute of Science Education and Research (IISER), Berhampur 760010 , India}
\affiliation{Indian Institute of Science Education and Research (IISER) Tirupati, Tirupati 517507, India}
\affiliation{Indian Institute Technology, Patna, Bihar 801106, India}
\affiliation{Indiana University, Bloomington, Indiana 47408}
\affiliation{Institute of Modern Physics, Chinese Academy of Sciences, Lanzhou, Gansu 730000 }
\affiliation{University of Jammu, Jammu 180001, India}
\affiliation{Joint Institute for Nuclear Research, Dubna 141 980, Russia}
\affiliation{Kent State University, Kent, Ohio 44242}
\affiliation{University of Kentucky, Lexington, Kentucky 40506-0055}
\affiliation{Lawrence Berkeley National Laboratory, Berkeley, California 94720}
\affiliation{Lehigh University, Bethlehem, Pennsylvania 18015}
\affiliation{Max-Planck-Institut f\"ur Physik, Munich 80805, Germany}
\affiliation{Michigan State University, East Lansing, Michigan 48824}
\affiliation{National Research Nuclear University MEPhI, Moscow 115409, Russia}
\affiliation{National Institute of Science Education and Research, HBNI, Jatni 752050, India}
\affiliation{National Cheng Kung University, Tainan 70101 }
\affiliation{Nuclear Physics Institute of the CAS, Rez 250 68, Czech Republic}
\affiliation{Ohio State University, Columbus, Ohio 43210}
\affiliation{Institute of Nuclear Physics PAN, Cracow 31-342, Poland}
\affiliation{Panjab University, Chandigarh 160014, India}
\affiliation{Pennsylvania State University, University Park, Pennsylvania 16802}
\affiliation{NRC "Kurchatov Institute", Institute of High Energy Physics, Protvino 142281, Russia}
\affiliation{Purdue University, West Lafayette, Indiana 47907}
\affiliation{Rice University, Houston, Texas 77251}
\affiliation{Rutgers University, Piscataway, New Jersey 08854}
\affiliation{Universidade de S\~ao Paulo, S\~ao Paulo, Brazil 05314-970}
\affiliation{University of Science and Technology of China, Hefei, Anhui 230026}
\affiliation{Shandong University, Qingdao, Shandong 266237}
\affiliation{Shanghai Institute of Applied Physics, Chinese Academy of Sciences, Shanghai 201800}
\affiliation{Southern Connecticut State University, New Haven, Connecticut 06515}
\affiliation{State University of New York, Stony Brook, New York 11794}
\affiliation{Temple University, Philadelphia, Pennsylvania 19122}
\affiliation{Texas A\&M University, College Station, Texas 77843}
\affiliation{University of Texas, Austin, Texas 78712}
\affiliation{Tsinghua University, Beijing 100084}
\affiliation{University of Tsukuba, Tsukuba, Ibaraki 305-8571, Japan}
\affiliation{United States Naval Academy, Annapolis, Maryland 21402}
\affiliation{Valparaiso University, Valparaiso, Indiana 46383}
\affiliation{Variable Energy Cyclotron Centre, Kolkata 700064, India}
\affiliation{Warsaw University of Technology, Warsaw 00-661, Poland}
\affiliation{Wayne State University, Detroit, Michigan 48201}
\affiliation{Yale University, New Haven, Connecticut 06520}

\author{J.~Adam}\affiliation{Brookhaven National Laboratory, Upton, New York 11973}
\author{L.~Adamczyk}\affiliation{AGH University of Science and Technology, FPACS, Cracow 30-059, Poland}
\author{J.~R.~Adams}\affiliation{Ohio State University, Columbus, Ohio 43210}
\author{J.~K.~Adkins}\affiliation{University of Kentucky, Lexington, Kentucky 40506-0055}
\author{G.~Agakishiev}\affiliation{Joint Institute for Nuclear Research, Dubna 141 980, Russia}
\author{M.~M.~Aggarwal}\affiliation{Panjab University, Chandigarh 160014, India}
\author{Z.~Ahammed}\affiliation{Variable Energy Cyclotron Centre, Kolkata 700064, India}
\author{I.~Alekseev}\affiliation{Alikhanov Institute for Theoretical and Experimental Physics NRC "Kurchatov Institute", Moscow 117218, Russia}\affiliation{National Research Nuclear University MEPhI, Moscow 115409, Russia}
\author{D.~M.~Anderson}\affiliation{Texas A\&M University, College Station, Texas 77843}
\author{A.~Aparin}\affiliation{Joint Institute for Nuclear Research, Dubna 141 980, Russia}
% ...

\collaboration{STAR Collaboration}\noaffiliation

\date{\today}

\begin{abstract}
Open heavy flavor hadrons provide unique probes of the medium produced in ultra-relativistic heavy-ion collisions. Due to their increased mass relative to light-flavor hadrons, long lifetime, and early production in hard-scattering interactions, they provide access to the full evolution of the partonic medium formed in heavy-ion collisions. This paper reports two-dimensional (2D) angular correlations between neutral $D$-mesons and unidentified charged particles produced in minimum-bias Au+Au collisions at $\sqrt{s_{\rm NN}}$ = 200 GeV. $D^0$ and $\bar{D}^0$ mesons are reconstructed via their weak decay to $K^{\mp} \pi^{\pm}$ using the Heavy Flavor Tracker (HFT) in the Solenoidal Tracker at RHIC (STAR) experiment. Correlations on relative pseudorapidity and azimuth $(\Delta\eta,\Delta\phi)$ are presented for peripheral, mid-central and central collisions with $D^0$ transverse momentum from 2 to 10~GeV/$c$. Attention is focused on the 2D peaked correlation structure near the triggered $D^0$-meson, the {\em near-side} (NS) peak, which serves as a proxy for a charm-quark containing jet. The correlated NS yield of charged particles per $D^0$-meson and the 2D widths of the NS peak increase significantly from peripheral to central collisions. These results are compared with similar correlations using unidentified charged particles, consisting primarily of light-flavor hadrons, at similar trigger particle momenta. Similar per-trigger yields and widths of the NS correlation peak are observed. The present results provide additional evidence that $D^0$-mesons undergo significant interactions with the medium formed in heavy-ion collision and show, for the first time, significant centrality evolution of the NS 2D peak in the correlations of particles associated with a heavy-flavor hadron produced in these collisions.
\end{abstract}


\maketitle

\section{Introduction}
\label{SecI}
Heavy-flavor (HF) quark (charm and beauty) production in relativistic heavy-ion collisions provides a unique probe of the produced deconfined partonic matter. This is because the energy scale (3~GeV for charm production) is sufficiently large such that the production mechanisms can be calculated from perturbative QCD (e.g. FONLL~\cite{Vitev,Nahrgang,HFdissoc}). The charm quark contained in the final-state particle is very likely produced in the initial collision stages~\cite{Vitev,HFFragm}. The charm quark and/or charmed hadron can therefore access the many-body QCD dynamics in the very early collision stage when the partonic density is largest. This enables experimental studies of (1) heavy flavor quark or hadron interactions in the medium, (2) medium modifications of heavy flavor quark fragmentation~\cite{borghini} and hadronization, and (3) dissociation mechanisms of HF mesons with hidden flavor (e.g. J/$\psi$)~\cite{HFdissoc}. Experimentally, open HF mesons can be indirectly observed via semi-leptonic decay modes to single electrons, or directly via weak decay channels, e.g. D$^0 \rightarrow K^- \pi^+$ and $\bar{\rm D}^0 \rightarrow K^+ \pi^-$. The latter two decay channels are used in the present analysis, and D$^{0}$ will be used to represent both D$^{0}$ and $\bar{\rm D}^0$ throughout this paper.

In recent years, HF yields in the form of transverse momentum ($p_T$) spectra and nuclear modification factor $R_{\rm AA}$~\cite{D0RAASTAR,D0RAAALICE,D0RAACMS,RAA_ATLAS,RAA_PHENIX}, and azimuthal anisotropy amplitude $v_2$~\cite{RAA_ATLAS,RAA_PHENIX,D0v2STAR,D0v2ALICE,D0v2CMS} have been reported from relativistic heavy-ion collision experiments at the Relativistic Heavy Ion Collider (RHIC) and at the Large Hadron Collider (LHC). Additionally, HF correlations using D-mesons have been studied by the ALICE collaboration in p+p and p+Pb collisions, and the results have been shown to be consistent with {\sc PYTHIA} \cite{ALICEyields}. While overall charm quark production (c$\bar{\rm c}$) follows binary nucleon + nucleon collision scaling, the yields at $p_{T} > 2$~GeV/$c$ are suppressed in heavy-ion collisions, relative to binary scaling expectations as seen in the measurements of open-charm hadron $R_{\rm AA}$~\cite{D0RAASTAR,D0RAAALICE,D0RAACMS}. Comparable amounts of suppression are also observed for light flavor (LF) meson production. In addition, $D^0$-meson $v_2$, as a function of transverse kinetic energy, is also comparable to LF results and is consistent with the number of constituent quark scaling observed for LF hadrons~\cite{D0v2STAR,D0v2ALICE,D0v2CMS}. Both results suggest that the charm quark or meson interacts significantly with the medium. Understanding of charm-quark energy loss could be enhanced with an observable other than $R_{AA}$ and $v_{2}$, such as two-particle correlations with a heavy-flavor meson as a trigger. For HF quarks, QCD predicts less radiative energy loss than for low mass quarks (``dead cone effect'')~\cite{deadcone} due to the suppression of gluon radiation at forward angles, below $M_{quark}/E_{quark}$~\cite{Eloss}. Collisional energy loss scales with inverse mass~\cite{Eloss}, suggesting further reduction in interaction effects relative to LF mesons. The surprisingly strong charm quark interaction effects implied by measurements of open-charm $R_{\rm AA}$ and $v_2$ therefore motivate additional measurements and new studies to better understand these interactions, e.g. measurements of possible long-range correlations on pseudorapidity.

Transverse momentum integrated, two-dimensional (2D) angular correlations of unidentified charged-particles from minimum-bias Au+Au collisions at RHIC energies have been measured by the STAR experiment~\cite{axialCI}. These correlations exhibit a sudden onset, starting near mid-central collisions, of an increase in the per-trigger amplitude and width along relative pseudorapidity of the near-side (NS), 2D correlation peak. Similarly, this elongated correlation structure, commonly referred to as the ``ridge,''\footnote{In this paper, ``ridge" refers to a near-side, long-range correlation on relative pseudorapidity ($\Delta\eta$), other than a quadrupole, $\cos(2\Delta\phi)$.} has been reported in $p_T$-selected, trigger-associated correlations in Au+Au collisions~\cite{JoernRidge,KoljaPaper} and in Pb+Pb collisions at the LHC~\cite{AlicePbPbRidge,AtlasPbPbRidge,CMSPbPbRidge}. The primary goals of the present analysis are to study the centrality dependence of 2D angular correlations of $D^{0}$-mesons plus associated charged hadrons, and to determine if the ridge phenomenon also occurs for HF mesons. 

In general, the HF $R_{\rm AA}$ and $v_2$ data from RHIC and the LHC have been described by a variety of models. The principle physics issues considered in the models include: (1) description of the initial-state including shadowing and saturation~\cite{glasma}; (2) the rapid HF formation time restricting the HF parton cascade~\cite{cascade}; (3) radiative, collisional (diffusion), and dissociative interactions~\cite{HFdissoc,Eloss}; (4) longitudinal color-field (glasma) effects on HF fragmentation~\cite{HF-CGC}; (5) HF hadronization based on fragmentation, recombination or a mixture of both~\cite{Li,CaoPRC92,Song,CaoPRC94,He}. Transport models~\cite{CaoPRC94,Song,Uphoff,Nahrgang} or stochastic transport of the HF quark or hadron within a hydrodynamic medium~\cite{Prado,CaoPRC92,He} are typically assumed. By reporting an experimental quantity, other than $R_{\rm AA}$ and $v_2$, which gives access to charm-jet and flow-related physics simultaneously, more sensitivity to the many-body, non-perturbative QCD interactions is possible. Future theoretical analyses of the data presented here may lead to a better understanding of those interactions and of the medium itself.

In the present work we report 2D angular correlations on relative pseudorapidity $\Delta\eta = \eta_{D^{0}} - \eta_{h^{\pm}}$ and relative azimuthal angle $\Delta\phi = \phi_{D^{0}} - \phi_{h^{\pm}}$ of charged-hadrons with $D^0$ and $\bar{D}^0$ mesons produced in minimum-bias Au+Au collisions at $\sqrt{s_{\rm NN}}$ = 200~GeV. This type of analysis permits the $(\Delta\eta,\Delta\phi)$ dependences of the correlations to be separated, allowing any possible $\Delta\eta$-dependent correlations to be distinguished from the $\Delta\eta$-independent azimuthal harmonics, such as elliptic flow. In this paper, we focus on the centrality evolution of the 2D angular distribution and number  of associated charged hadrons on the NS within $|\Delta\phi| \leq \pi/2$ which have become correlated with the triggered $D^0$-meson as the charm quark fragments or the $D^0$-meson propagates through the medium. 

This paper is organized as follows. The analysis method is described in Sec.~\ref{SecII} and the data processing steps and other details are discussed in Sec.~\ref{SecIII} and in two appendices. The correlation data and fitting results are presented in Sec.~\ref{SecIV}. Systematic uncertainties are discussed in Sec.~\ref{SecV} and our results are further discussed and compared with predictions from {\sc PYTHIA} and LF correlations measured by STAR in Sec.~\ref{SecVI}.  A summary and conclusions from this work are given in Sec.~\ref{SecVII}.

\section{Analysis Method}
\label{SecII}
In conventional two-particle correlation analyses with charged particles, the number of particle pairs from the same-event (SE) in each $(\Delta\eta,\Delta\phi)$ bin is summed for a collection of collision events (e.g. in a centrality class). The particle pairs are primarily uncorrelated, but do include correlation effects. The uncorrelated pair background can be estimated by similarly counting pairs of particles where the two particles in each pair are sampled from different events (mixed-events, ME). The mixed-events are required to have similar multiplicities and primary collision vertex positions along the beam axis. Mixed-events were not selected based on event-plane orientation so that the measured correlations include the contributions from the azimuthal anisotropy of the particle distributions. These quantities are defined as $\rho_{\rm SE}(\Delta\eta,\Delta\phi)$ and $\rho_{\rm ME}(\Delta\eta,\Delta\phi)$, respectively, which are pair densities (number of pairs per $(\Delta\eta,\Delta\phi)$ bin area). The normalized difference, $\Delta\rho \equiv \rho_{\rm SE} - \alpha\rho_{\rm ME}$, approximates the two-particle correlation in each $(\Delta\eta,\Delta\phi)$ bin, where $\alpha$ is the ratio of the total number of SE pairs to ME pairs, $\alpha \equiv N_{\rm pairs,SE}/N_{\rm pairs,ME}$, which normalizes $\rho_{\rm ME}$ to the same overall scale as $\rho_{\rm SE}$. Both pair counts are affected by detector acceptance and particle reconstruction inefficiency. These effects can be corrected in each $(\Delta\eta,\Delta\phi)$ bin by dividing $\Delta\rho$ by the normalized ME distribution, $\alpha\rho_{\rm ME}$. The ratio $\Delta\rho/\alpha\rho_{\rm ME}(\Delta\eta,\Delta\phi)$ [see Eq.~(1) in Ref.~\cite{axialCI}] is the underlying acceptance and efficiency corrected correlation quantity from which other quantities, e.g. per-trigger correlations, can be derived. 

For $D^0$-meson plus charged-particle ($D^0 + h^{\pm}$) correlations, where the short-lived $D^0$ cannot be detected directly, only statistical reconstruction is possible because the number of produced $D^0$-mesons must be inferred from the invariant mass distribution constructed from the daughter particle momentum vectors. The cleanest decay channel for $D^{0}$ reconstruction is the weak decay to unlike-sign $K^{\mp}\pi^{\pm}$ pairs (B.R. = 3.93\%)~\cite{PDG}. Random, combinatoric $K^{\mp}\pi^{\pm}$ pairs which pass all the track and vertex reconstruction cuts can be drastically reduced based on the optimized secondary decay vertex parameters. These parameters are evaluated using particle trajectories (tracks) measured in the STAR Time Projection Chamber (TPC)~\cite{TPC} and Heavy Flavor Tracker (HFT)~\cite{HFTCDR}.

Even so, background pairs remain and it is impossible to distinguish random $K^{\mp}\pi^{\pm}$ pairs from true $D^0$ decay daughters. Correlations between those random $K\pi$ pairs and other charged hadrons must be accounted for and removed in the analysis. These background correlations can be estimated by correlating random $K\pi$ pairs from side-bands (SB) in the $K\pi$ invariant mass distribution, with other charged particles in the event, as further explained in Sec.~\ref{SecIII}. Unless stated explicitly, both $K^{\mp}\pi^{\pm}$ pairs are included and denoted simply as $K\pi$.

In addition, a significant fraction of the $D^{0}$-mesons are produced from decays of the $D^{\star \pm}$ resonance. This fraction is estimated from the charm-quark fragmentation fractions into direct-$D^{0}$ (0.200), $D^{\star 0}$ (0.213) and $D^{\star \pm}$ (0.224) (see Sec. 17.8.1 in Ref. [33]) and the corresponding branching ratios (BR) for $D^{*0} \rightarrow D^{0} + \pi^{0}$ or $\gamma$ (BR = 100\%) and $D^{\star \pm} \rightarrow D^{0} + \pi^{\pm}$ (BR = 67.7\%). The resulting fraction of charm quarks which produce a $D^{0}$ is 0.565, of which 0.152 were from $D^{\star \pm}$ decays. We therefore estimate that approximately 27\% of the $D^{0}$ sample is from $D^{\star \pm}$ decays.

Correlations between these daughter $D^0$-mesons from $D^{\star \pm}$ decays and other charged particles are of physical interest because the daughter $D^0$-meson contains the original c-quark and most of the parent $D^{\star \pm}$ momentum. However, the decay length of the $D^{\star \pm}$ ($\sim$0.12 nm) indicates that the $D^{\star \pm} \rightarrow D^0 + \pi^{\pm}$ decay happens well outside the medium, producing an additional low-momentum, {\em soft} charged pion ($\pi_{\rm soft}$). The resulting $D^0 + \pi_{\rm soft}$ correlation reflects only the decay kinematics and is considered a background correlation which must also be removed from the measurements.

The true $D^0 + h^{\pm}$ correlations are calculated by subtracting the random $K\pi + h^{\pm}$ background correlations and the preceding correlated $D^0 + \pi_{\rm soft}$ pairs from the measured quantity $\Delta\rho/\alpha\rho_{\rm ME}$, where the latter uses all $K\pi$ pairs in the $D^0$ signal range of the $K\pi$ invariant mass distribution. The basic correlation quantity, derived in Appendix~A, is given by
\bea
\frac{\Delta\rho_{D^0+h}}{\alpha \rho_{{\rm ME},D^0+h}}  & = & 
\frac{S+B}{S} \frac{\Delta\rho_{\rm sig} - \Delta\rho_{D^0 + \pi_{\rm soft}}}
                   {\alpha \rho_{\rm ME,sig}}
\nonumber \\
  & - & \frac{B}{S} \frac{\Delta\rho_{\rm SB}}{\alpha_{\rm SB} \rho_{\rm ME,SB}}
\label{Eq1}
\eea
where $S$ and $B$ are the deduced $D^0$ signal and background yields from the $K\pi$ invariant mass distribution near the $D^0$ mass, as explained in the next section. Subscripts $D^0+h$, $D^0 + \pi_{\rm soft}$, sig, and SB indicate the pair and $K\pi$ invariant mass region used. The quantity on the left-hand side of Eq.~(\ref{Eq1}) is symbolic only and must be determined by the measured quantities on the right-hand side.  Other technical details for these calculations are explained in the next section.

\section{Data and technical details}
\label{SecIII}

\subsection{Event and particle selection}

The data for this analysis were collected by the STAR experiment~\cite{STARNIM} at RHIC during the 2014 run period. The dataset includes approximately 900 million minimum-bias Au+Au $\sqrt{s_{NN}} = $ 200~GeV collision events for which coincident signals between the two, symmetrically positioned Vertex Position Detectors (VPD)~\cite{VPDStarPaper2014} were required. In addition, the primary collision vertex (PV) for each accepted event was required to be within the tracking fiducial region of the HFT~\cite{HFTCDR}, $\pm$6 cm along the beam axis ($z$-axis), to ensure uniform HFT acceptance.

Charged particle trajectories were initially reconstructed using the TPC in the presence of a 0.5 T uniform magnetic field parallel with the beam axis. All used tracks were required to have at least 20 reconstructed space points (``hits'') (out of a possible 45) in the TPC, a ratio of the number of found hits to the maximum number expected $>$ 0.52 to remove split tracks, a Kalman filter least-squares fitted $\chi^{2}/{\rm NDF} < 3$, and a distance of closest approach (DCA) to the PV $<$~3~cm. All tracks used in the analysis must also fall within the acceptance of the STAR TPC: $p_T> 0.15$~GeV/$c$, $|\eta| \leq 1$ (pseudorapidity), and full $2\pi$ in azimuth. This sub-set of reconstructed charged particles in the TPC are referred to as {\em TPC tracks} and were used to determine collision centrality (see next sub-section). In addition, all tracks used to construct the correlations were required to match to at least one hit in each of the inner three layers of the HFT~\cite{HFTCDR}, including two in the silicon pixel detector (PXL) layers, and one in the Intermediate Silicon Tracker (IST)~\cite{IST}. The spatial resolution of projected tracks near the PV is greatly improved, e.g. from $\sim$~1~cm for TPC tracks to $\sim 30$~$\mu$m, for $p_{T} > 1.5$ GeV/$c$, when HFT hits are included. Furthermore, the fast-timing of the IST suppresses track pileup contamination from collisions occurring before or after the triggered collision.  

\begin{figure}[t]
\includegraphics[keepaspectratio,width=3.3in]{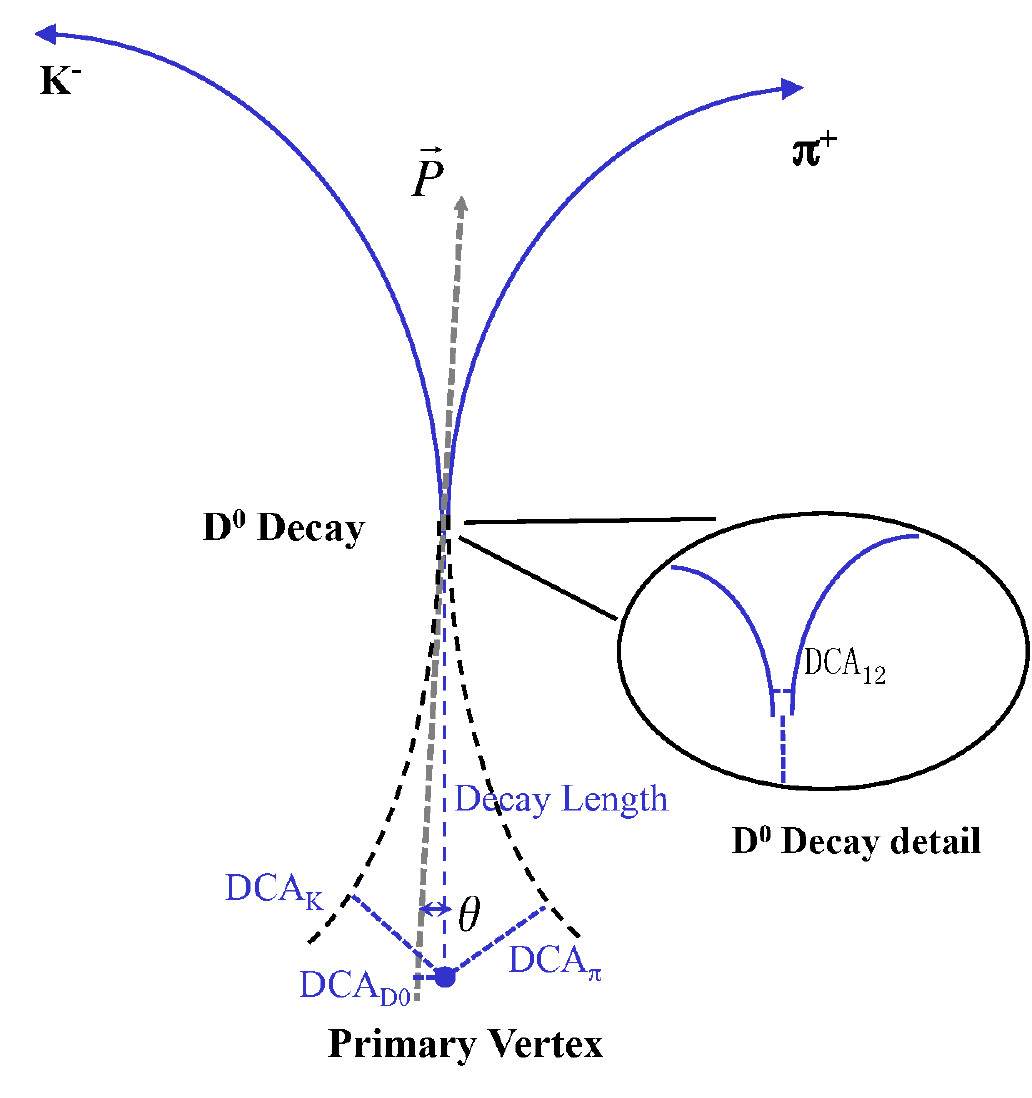}
\centering
\caption{$D^{0}$ decay diagram depicting the five topological reconstruction cuts~\cite{D0RAASTAR}: 1) decay length, 2) DCA between decay daughters ($\rm{DCA}_{12}$), 3) DCA of reconstructed $D^{0}$ to PV, where $\theta$  (the angle between the $D^{0}$ momentum vector and the straight line between the primary and $D^{0}$ decay vertices) and the decay length are used in the calculation, 4) DCA of daughter pion to PV, and 5) DCA of daughter kaon to PV. }
\centering
\label{d0DecayFigure}
\end{figure}

The trigger $D^{0}$ and $\bar{D}^0$ were reconstructed via the hadronic decay channel $D^{0} \rightarrow K+\pi$ ($c\tau$ = 123~$\mu$m for $D^{0}$). The secondary decay vertices are reconstructed using the above TPC+HFT tracks and an optimized set of limits on the accepted decay topology parameters (see Fig. \ref{d0DecayFigure}). The parameter limits were taken from a previous STAR $D^{0}$ analysis~\cite{D0v2STAR} after adjusting for the 2 $<$ $p_{T,D^{0} }$ $<$ 10 GeV/$c$ range used in the present analysis. The limits for the five geometrical cuts in Fig.~\ref{d0DecayFigure} $-$ the decay length, DCA between daughters, DCA of the reconstructed $D^{0}$ to the PV, DCA of the pion daughter to the PV, and DCA of kaon daughter to the PV are, respectively, $>212$~$\mu$m, $<57$~$\mu$m, $<38$~$\mu$m, $>86$~$\mu$m, and $>95$~$\mu$m. In addition, pion and kaon identification requirements based on measured ionization energy loss ($dE/dx$) in the TPC were imposed on the $D^{0}$ decay daughter candidates. Those cuts required that the fitted $dE/dx$ for the assigned space-points be $<2 \sigma$ from the expected mean. 

\begin{figure*}[t]
\includegraphics[keepaspectratio,width=6.5in]{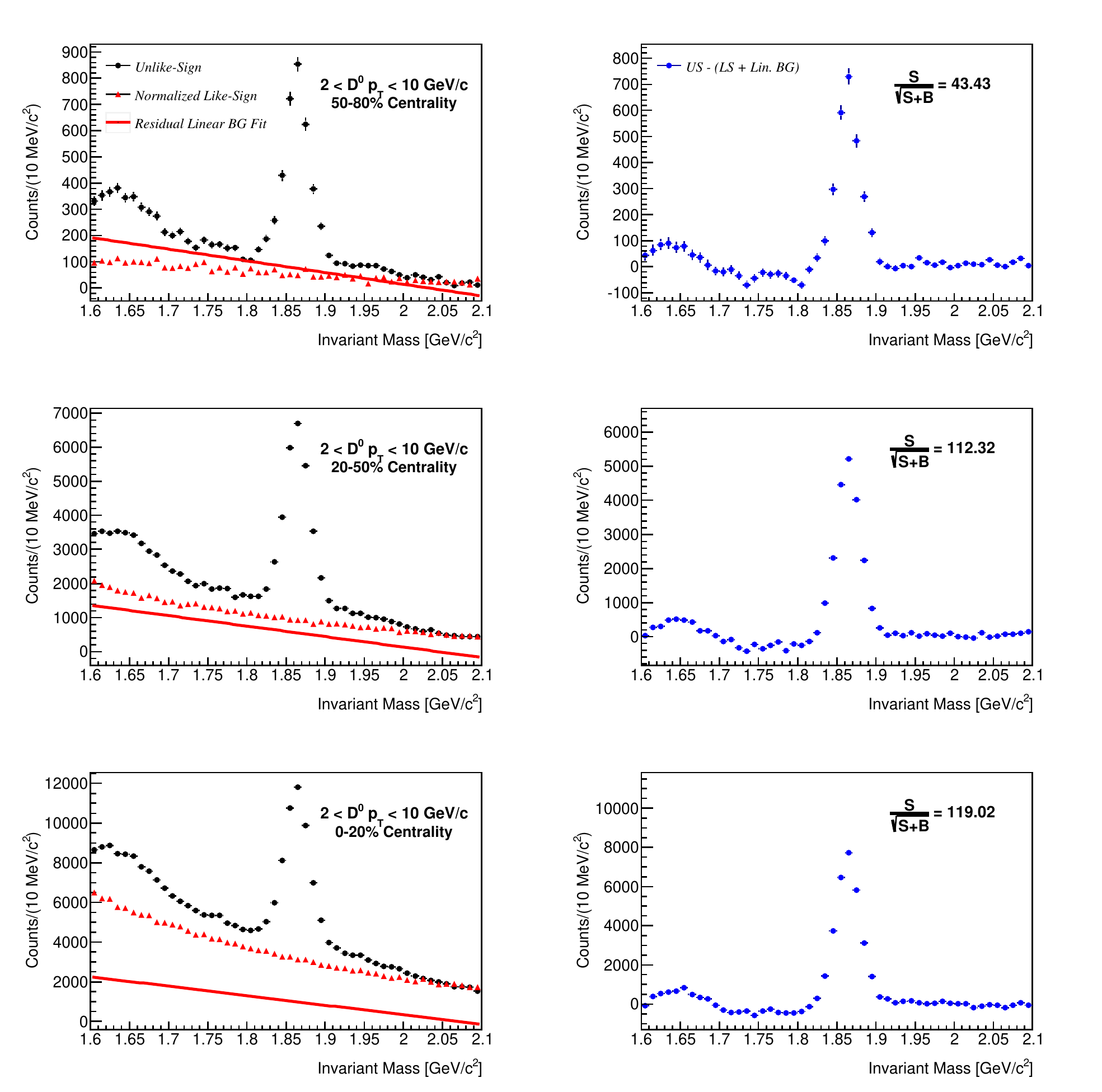}
\centering
\caption{Invariant mass distributions for the three centrality classes used in this analysis from peripheral to central in the upper to lower rows of panels, respectively. The US (dots) and normalized LS (triangles) distributions are shown in the left-hand column, and the LS + linear background subtracted distributions are shown in the right-hand column. The linear background fit functions for the residual background after subtraction of the normalized LS distributions are shown by the straight lines in the left-hand panels. }
\centering
\label{invariantMassDistributions}
\end{figure*}

The invariant mass distribution was constructed for all unlike-sign (US) and like-sign (LS) $K\pi$ pairs, where the LS approximates the invariant mass background. The LS distribution was normalized to the US data in the invariant mass range from $2.0 < M_{K\pi} < 2.1$~GeV/$c^{2}$ and subtracted from the US distribution. The remaining background was fit with a linear function and then subtracted. Other functional shapes (e.g. exponential, polynomial) and fitting ranges were tested and the resulting variations in the signal (S) and background (B) yields were found to be small ($<$ 1\%). The systematic effect on the correlations due to the extraction of the S and B yields are discussed in ~Sec.\ref{SecV}. The results of this procedure are summarized in Fig. \ref{invariantMassDistributions}. The signal and background yields were calculated using bin counting in the invariant mass distribution in the range $1.82 < M_{K\pi} < 1.90$~GeV/$c^{2}$ ($\pm$2$\sigma$), where S+B was calculated using the raw, unlike-sign distribution, and S was calculated from the fully-subtracted distribution. All trigger $D^{0}$s used in the present analysis were restricted within $2 < p_{T,D^{0}} < 10$~GeV/$c$ to maximize statistical significance. Correlations constructed with $p_{T,D^{0}} < 2$~GeV/$c$ exhibited large fluctuations in the correlation structures for small changes in the topological cuts, far beyond statistical uncertainty, and were therefore excluded from the analysis. This instability is likely due to the significant increase of mismatching between TPC tracks and HFT hits at low $p_{T}$, resulting in a smaller signal-to-background ratio for $D^{0}$ below 2 GeV/$c$. Residual structure in the invariant mass background below 1.7 GeV/$c^2$ was found to be predominately from other $D^{0}$-meson decays \cite{D0v2STAR}.

\subsection{Collision centrality determination}

The minimum-bias event sample was divided into three centrality classes, using the observed event-wise number of TPC tracks with $|\eta| \leq 1$ and $p_T \geq 0.15$~GeV/$c$ according to the method in Ref.~\cite{axialCI}. The measured multiplicity frequency distribution was corrected for TPC tracking efficiency, thus determining TPC track multiplicity limits corresponding to centrality fractions 0-20\%, 20-50\% and 50-80\% of the total reaction cross section. Centrality was based on multiplicities within $|\eta| \leq 1$, instead of $|\eta| \leq 0.5$, in order to avoid significant artifacts in the angular correlations along $\Delta\eta$ as explained in Ref.~\cite{axialCI}. Additional corrections due to small (few percent) variations in TPC tracking efficiency as functions of PV position and beam + beam run-time luminosity were negligible. 

\subsection{Construction of pair histograms}

$D^{0}$-candidate + associated charged-hadron pairs from the same event are formed on binned coordinates ($\Delta\eta, \Delta\phi$) and summed over all events in each centrality class. Particles used as $D^{0}$ daughter-candidates are excluded from the associated hadrons. The $D^{0}$ trigger is defined to be any reconstructed $K\pi$ pair passing all the above cuts and falling within the invariant mass range $1.82 < M_{K\pi} < 1.90$~GeV/$c^{2}$. This selection includes both real $D^{0}$s as well as combinatorial background $K\pi$ pairs. To estimate the correlations from this background, pair histograms on ($\Delta\eta, \Delta\phi$) are also constructed using $K\pi$ pairs from two side-band regions in the invariant mass spectrum defined by: left side-band, $1.7 < M_{K\pi} < 1.8$~GeV/$c^{2}$; right side-band, $1.92 < M_{K\pi} < 2.10$~GeV/$c^{2}$. The different widths of the left and right side-bands are chosen to use approximately the same yield of background $K\pi$ pairs from each side-band.  An efficiency correction (weight) is applied to each individual $K\pi$ + associated hadron pair. The pair-weight is defined as
\begin{equation}
pair \ weight = \frac{B}{S+B}\frac{\overline{\epsilon_{K}\epsilon_{\pi}\epsilon_{h}}}{\epsilon_{K}\epsilon_{\pi}\epsilon_{h}} +
                \frac{S}{S+B}\frac{\overline{\epsilon_{D^{0}}\epsilon_{h}}}{\epsilon_{D^{0}}\epsilon_{h}}
\label{efficiencyPairWeight}
\end{equation}
where $\epsilon_{D^{0}}$, $\epsilon_{K}$, $\epsilon_{\pi}$ and $\epsilon_{h}$ are the individual reconstruction efficiencies for the $D^{0}$, $K$, $\pi$ and charged-hadron, respectively. Overbars in Eq.(2) denote averages over all events in respective $p_{T,D^{0}}$ and centrality bins. Ratios $S/(S+B)$ and $B/(S+B)$ are the probabilities that the candidate $K\pi$ pair is actually from a $D^{0}$ decay or is random, respectively. In the side-bands all $K\pi$ pairs are considered random. The $K$, $\pi$ and charged-hadron TPC tracking efficiencies are taken from the analysis in Ref.~\cite{STARSpectraPaper}. Those efficiencies are then multiplied by the additional $p_{T}$-dependent, TPC+HFT track matching ratio to get the quantities $\epsilon_{K}$, $\epsilon_{\pi}$ and $\epsilon_{h}$. The $D^{0}$ efficiency is the ratio of the raw yield of $D^{0}$ mesons as a function of $p_{T}$, using the above cuts, to the published invariant yield~\cite{D0RAASTAR}.

The overall shape of the above $K\pi$ + hadron pair distribution is dominated by the finite pseudorapidity acceptance which introduces an approximate triangular shape on $\Delta\eta$. This overall shape plus any other acceptance artifacts caused by the TPC sector edges, electronics outages, etc.,~ can be corrected by dividing, bin-by-bin, a similarly constructed mixed-event distribution as explained in Sec.~\ref{SecII}. Accurate acceptance corrections require that the PV location of each pair of mixed-events are sampled within sufficiently narrow sub-bins along the beam-axis, where the 12~cm range was divided into 10 uniform sub-bins. Correlation artifacts can also occur when the centralities of the mixed-events differ too much. Restricting the range of multiplicties in the event-mixing sub-bins to $<$ 50, assuming 2 units in $\eta$, was previously shown to be sufficient~\cite{AyaCD,AyaCI,axialCI}. The latter resulted in 16 multiplicity sub-bins, for a total of 160 event-mixing sub-bins in this analysis. Mixed-event distributions were constructed for each of the three $K\pi$ invariant mass ranges discussed above. Efficiency corrections were also applied to each mixed-event $K\pi$ + hadron pair as in Eq.~(\ref{efficiencyPairWeight}). The $D^{0}$ reconstruction efficiency increases steeply from 2-5 GeV/$c$ by approximately a factor of 5-9, depending on the centrality bin, before reaching a plateau above 5 GeV/$c$. The correlations were also compared with and without the efficiency correction and the differences were negligible.

\subsection{Symmetrization on $\Delta\eta$ and $\Delta\phi$}

For the present analysis with identical, unpolarized colliding ions, where particles within a symmetric pseudorapidity range centered at $\eta = 0$ are used, we may project the above pair histograms onto absolute value binned coordinates $(|\Delta\eta|,|\Delta\phi|)$ without loss of information. Pair counts and statistical errors can then be copied to corresponding bins in the other quadrants in the full $(\Delta\eta,\Delta\phi)$ space for visual display. An odd number (13) of uniform $\Delta\eta$ bins within $-2 \leq \Delta\eta \leq 2$ were used and multiples of four $\Delta\phi$ bins (12) were assumed within full 2$\pi$, where $\Delta\phi$ bins are centered at $\Delta\phi = 0$ and $\pi$. Bins centered at $(\Delta\eta,\Delta\phi) = (0,0)$ and $(0,\pi)$ therefore have approximately 1/4 the number of pairs as nearby bins centered at non-zero $(\Delta\eta,\Delta\phi)$. Other bins centered at either $\Delta\eta = 0$, $\Delta\phi = 0$, or $\Delta\phi = \pi$ similarly have $\sim 1/2$ the number of pairs. Statistical errors in these bins are approximately factors of 2 and $\sqrt{2}$ larger, respectively, than that in neighboring non-zero $(\Delta\eta,\Delta\phi)$ bins. Statistical errors in the final correlations in Eq.~(\ref{Eq1}) are determined by the SE and ME pair counts in each $(|\Delta\eta|,|\Delta\phi|)$ bin, including the error contributions from the $D^0$ signal region and the two side-bands of the $K\pi$ invariant mass distribution, assuming uncorrelated uncertainties. Absolute statistical uncertainties in the correlations for each centrality class are approximately $\pm$0.0095 for 50-80\%, $\pm$0.004 for 20-50\%, and $\pm$0.002 for the 0-20\% centrality class. Errors generally increase by almost a factor of two at the outermost bins on $\Delta\eta$.

\subsection{$D^{\star \pm}$ correction} \label{dStarSubSec}

In Sec.~\ref{SecII}, the background contribution of $D^{\star \pm}$ decays occurring outside the medium to $D^{0}$+$\pi_{\rm soft}$ pairs was discussed. The charm-quark which forms the $D^{\star \pm}$ is created in an initial hard-scattering interaction and is therefore sensitive to the evolution of the medium. However, the $D^{\star \pm} \rightarrow D^{0} + \pi_{\rm soft}$ decay occurs outside the medium and the daughter $D^0$+soft-pion angular correlation is a result of vacuum decay. This contributes to the measured $D^0 + h^{\pm}$ correlation mainly in the $(\Delta\eta,\Delta\phi) = (0,0)$ bin. These decay daughter-pairs are treated as background. The number of such $D^{\star \pm}$ decays can be measured via a three-body invariant mass distribution constructed as $M_{K\pi\pi_{\rm soft}} - M_{K\pi}$ where the $D^{\star \pm}$ appears as a peak in the range [0.143,0.147]~GeV/$c^2$~\cite{Dstarpaper}. The $D^{\star \pm}$ yield and its background reference are normalized and used to correct the final correlations as described in Sec.~\ref{SecII} and in Eq.~(\ref{Eq1}). The peak amplitudes of the final correlations in the (0,0) bin are reduced by approximately 0.037 $\pm$ 0.012(stat.) $\pm$ 0.004(syst.), 0.046 $\pm$ 0.002(stat.) $\pm$ 0.018(syst.), and
0.015 $\pm$ 0.001(stat.) $\pm$ 0.013(syst.) in the 50-80\%, 20-50\% and 0-20\% centrality classes, respectively. The quoted statistical
errors are the statistical uncertainties in the correlation amplitude for $D^{0}$+hadron correlations coming from a $D^{*}$ decay. The systematic uncertainties are estimated from the variation in the deduced $D^{*}$ yields associated with 20\% changes in the magnitudes of the combinatoric background in the invariant mass distributions used to identify $D^{*}$ decays.  The total systematic effects from the $D^{*\pm}$ contamination and its correction are further discussed in Sec.~\ref{SecV}. 

\section{Results}
\label{SecIV}

\begin{figure*}[t]
\includegraphics[keepaspectratio,width=6.0in]{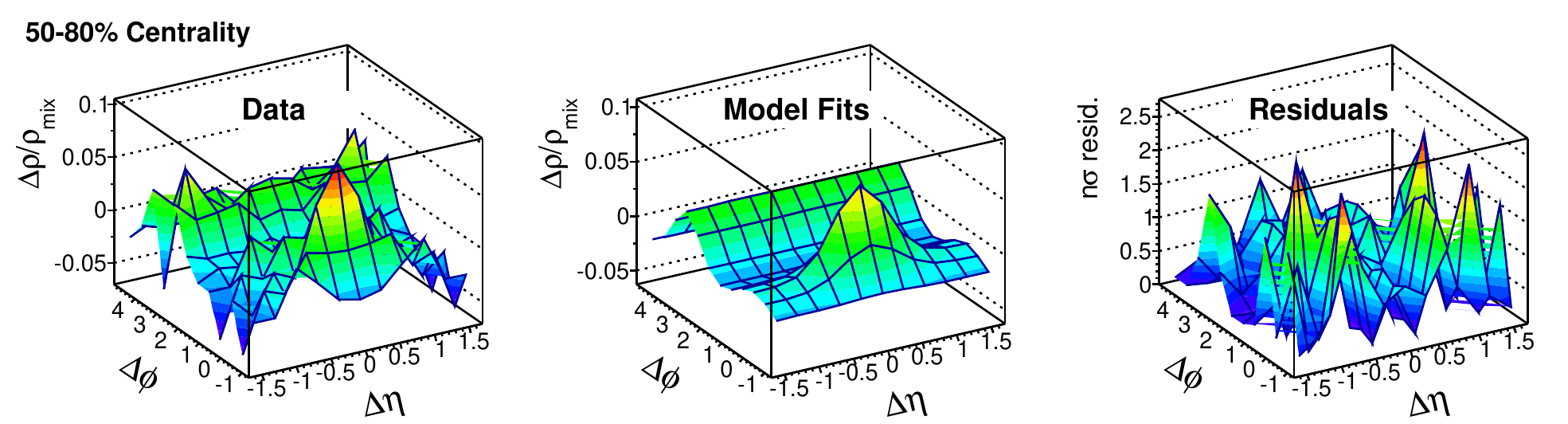}
\includegraphics[keepaspectratio,width=6.0in]{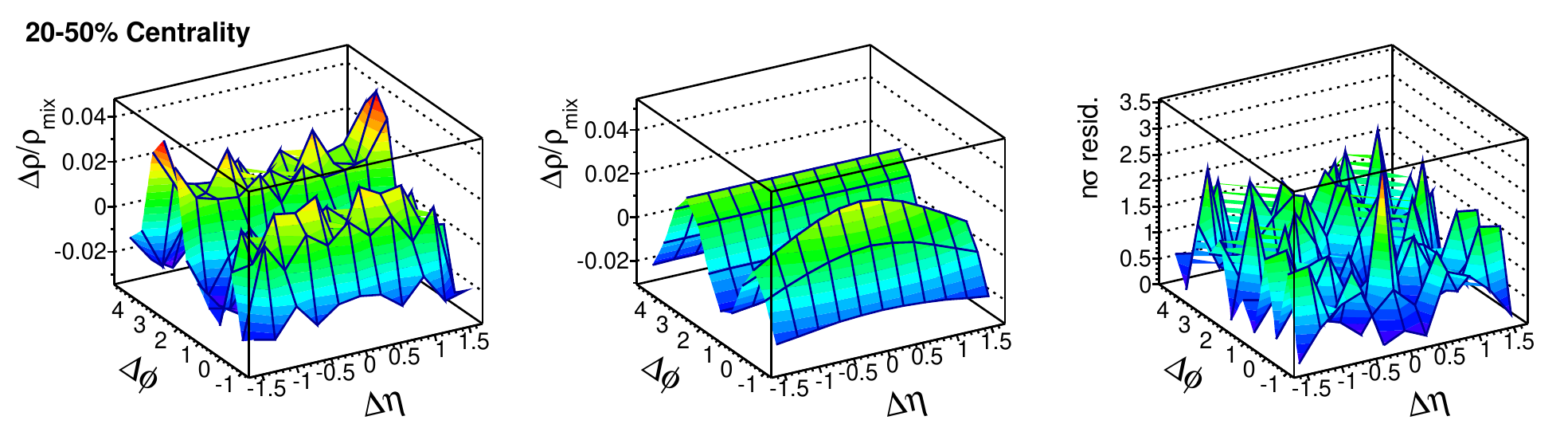}
\includegraphics[keepaspectratio,width=6.0in]{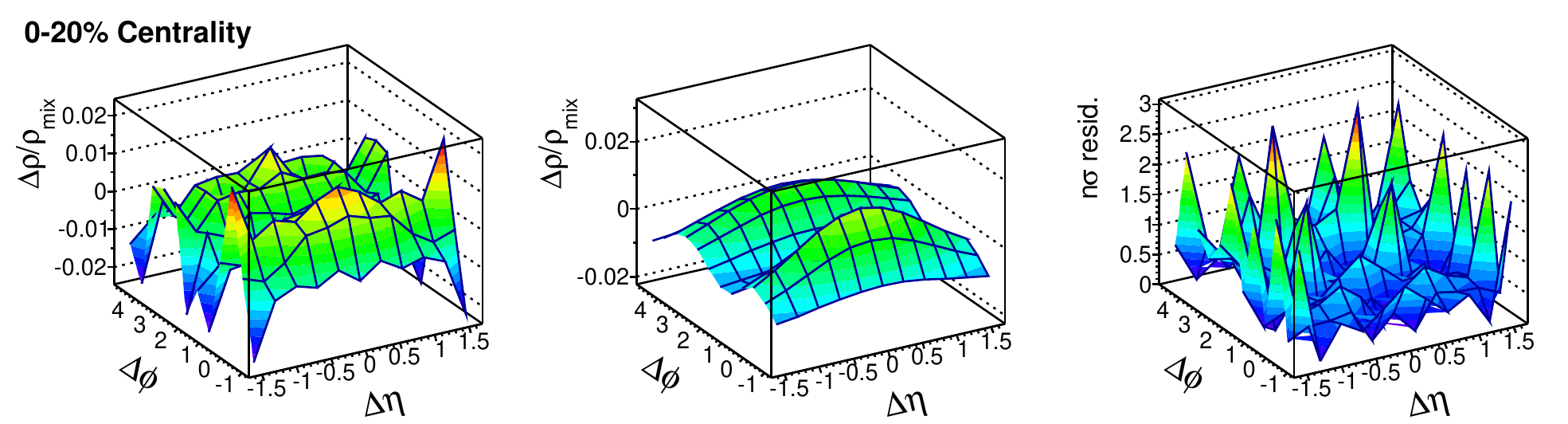}
\centering
\caption{\label{Fig1}
$D^0 + h^{\pm}$ correlation data (left column), model fits (middle column), and residuals in units of $|n\sigma|$ per bin (right column). Centrality fractions are 50-80\%, 20-50\%, and 0-20\% from top to bottom rows, respectively.}
\centering
\end{figure*}

The per-pair normalized $D^0 + h^{\pm}$ symmetrized correlations for centralities 50-80\%, 20-50\% and 0-20\% with 2 $<$ $p_{T,D^{0} }$ $<$ 10 GeV/$c$, defined in Eq.~(\ref{Eq1}), are shown in the left-hand column of panels in Fig.~\ref{Fig1}. Significant structures are visible in the correlations which exceed the statistical fluctuations, including $\Delta\eta$-independent near-side and away-side (AS) structures on $\Delta\phi$, and a near-side 2D peak which increases in width on $\Delta\eta$ with centrality. The structures are similar to the dominant features reported previously in unidentified and LF identified 2D dihadron angular correlations~\cite{axialCI,JoernRidge,KoljaPaper} from Au+Au collisions at 200~GeV. The broadening on $\Delta\eta$ of the NS 2D peak with increasing centrality, observed in these $D^0 + h^{\pm}$ correlations, is similar to that reported for unidentified dihadron correlations~\cite{axialCI}.

\begin{table*}[t]
\caption{Model parameters, statistical errors, and asymmetric systematic uncertainties (subscripts and superscripts) for $D^0$ + hadron correlations in Au + Au collisions at $\sqrt{s_{\rm NN}} =$ 200 GeV.}
\label{TableI}
\begin{tabular}{cccc}
\hline
Centrality(\%) & 50-80 & 20-50 & 0-20 \\
\hline \\[-1.5ex]
\vspace{0.02in}
$A_0$    & $-$0.012$\pm$0.004$^{+0.004}_{-0.002}$ & $-$0.009$\pm$0.001$^{+0.003}_{-0.003}$ & $-$0.012$\pm$0.003$^{+0.002}_{-0.002}$ \\
\vspace{0.02in}
$A_{\rm Q}$ & 0.004$\pm$0.003$^{+0.001}_{-0.002}$ & 0.0066$\pm$0.0030$^{+0.0004}_{-0.0003}$ & 0.0$\pm$0.0022$^{+0.0004}_{-0.0001}$ \\
\vspace{0.02in}
$A_{\rm NS}$ & 0.091$\pm$0.019$^{+0.008}_{-0.005}$ & 0.037$\pm$0.004$^{+0.010}_{-0.010}$ & 0.044$\pm$0.006$^{+0.004}_{-0.003}$ \\
\vspace{0.02in}
$\sigma_{\Delta\eta,NS}$ & 0.31$\pm$0.08$^{+0.01}_{-0.05}$ & 1.37$\pm$0.35$^{+0.29}_{-0.38}$ & 1.24$\pm$0.30$^{+0.34}_{-0.42}$ \\
\vspace{0.02in}
$\sigma_{\Delta\phi,NS}$ & 0.35$\pm$0.07$^{+0.04}_{-0.02}$ & 0.66$\pm$0.06$^{+0.06}_{-0.06}$ & 0.75$\pm$0.07$^{+0.08}_{-0.08}$ \\
\vspace{0.02in}
$A_{\rm AS}$ & 0.030$\pm$0.020$^{+0.009}_{-0.005}$ & - & - \\
\vspace{0.02in}
$\sigma_{\Delta\eta,AS}$ & - & - & 1.33$\pm$0.25$^{+0.34}_{-0.34}$ \\
\vspace{0.02in}
$\sigma_{\Delta\phi,AS}$ & 0.55$\pm$0.14$^{+0.11}_{-0.11}$ & - & - \\
\vspace{0.02in}
$A_{\rm D}$ & - & 0.016$\pm$0.012$^{+0.001}_{-0.0004}$ & 0.019$\pm$0.004$^{+0.003}_{-0.003}$ \\
$\chi^2$/DoF & 0.93  &  1.90  &  1.17 \\
\hline
\end{tabular}
\end{table*}

\begin{figure*}[t]
\includegraphics[keepaspectratio,width=2.2in]{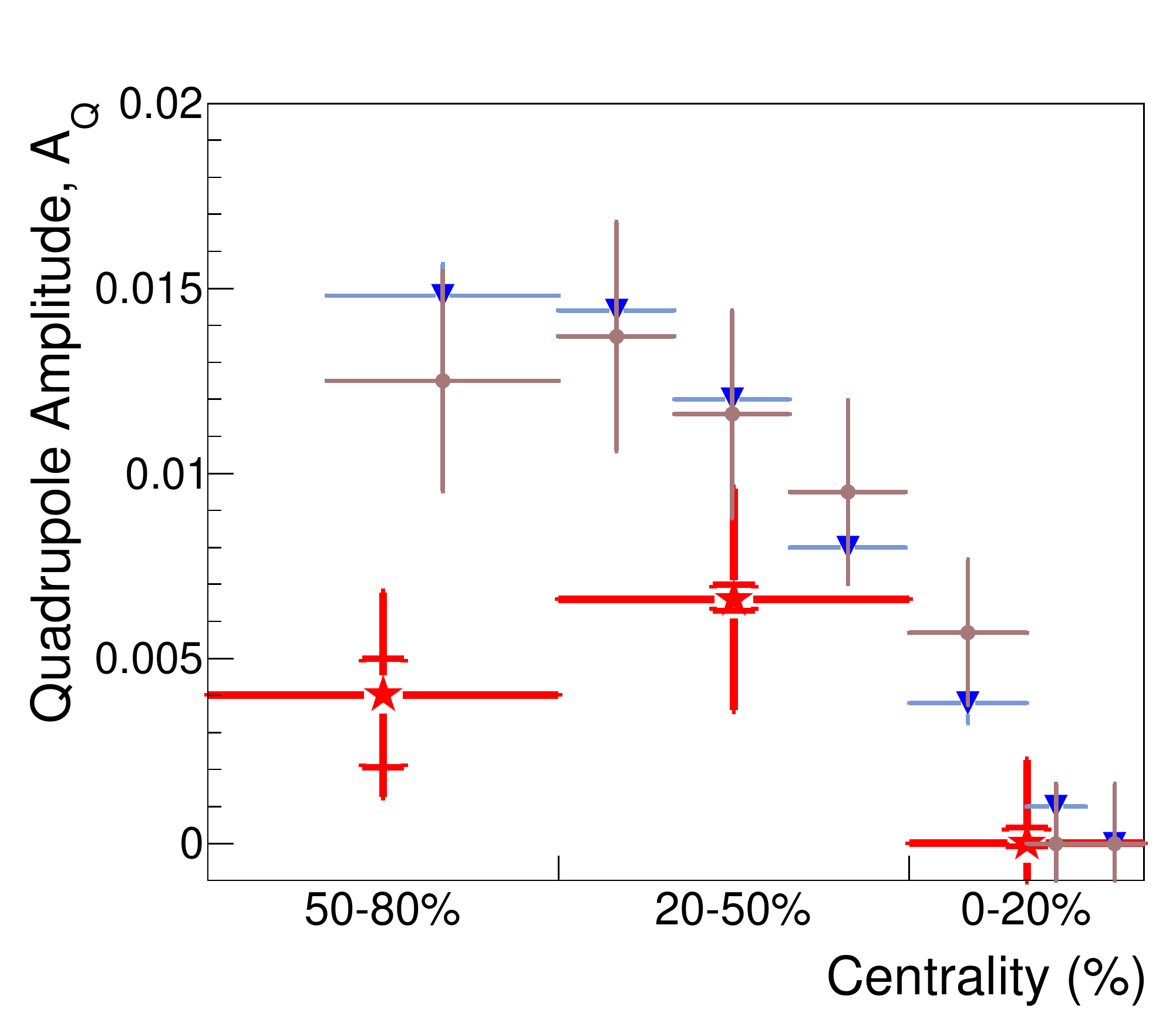}
\includegraphics[keepaspectratio,width=2.2in]{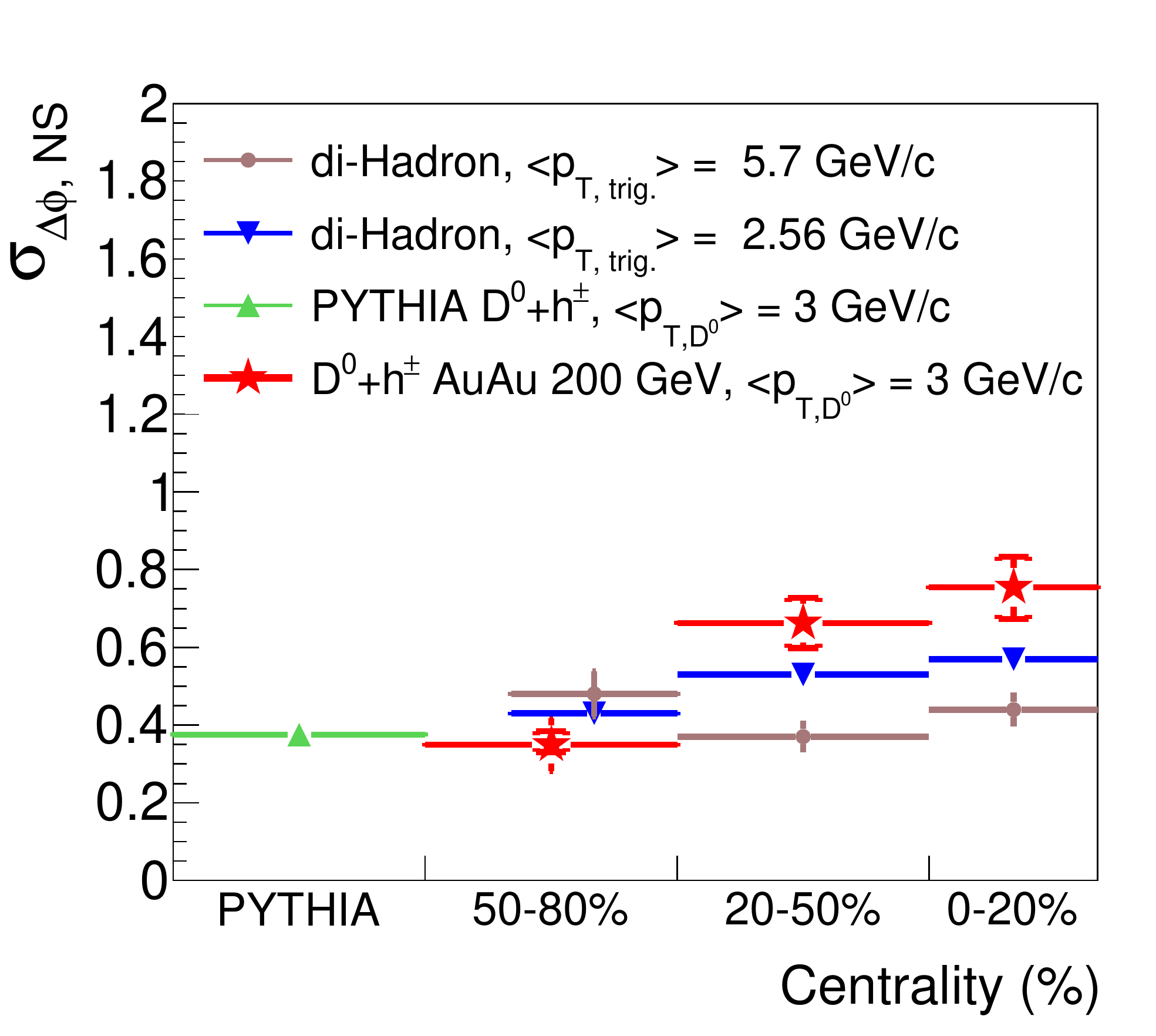}
\includegraphics[keepaspectratio,width=2.2in]{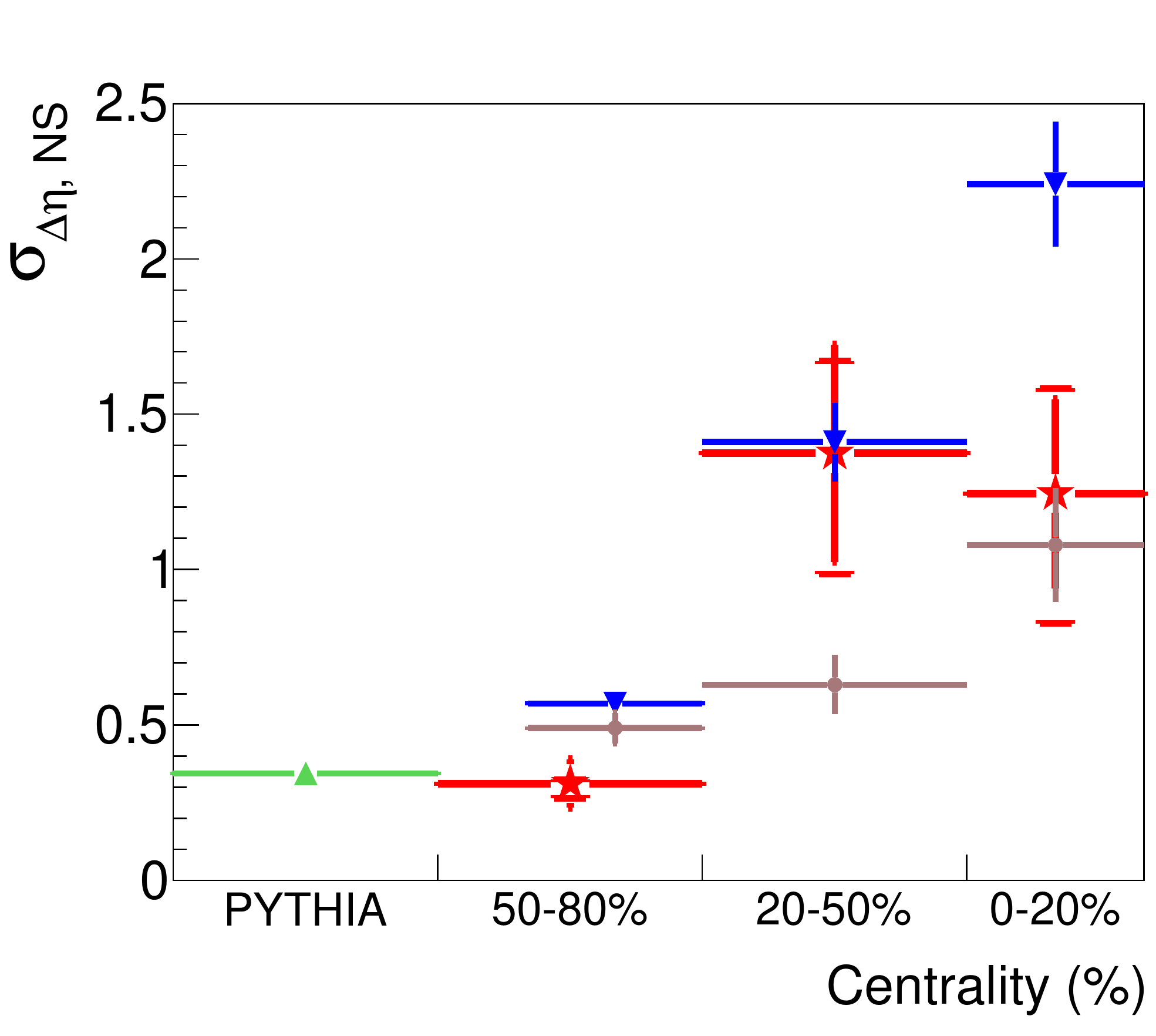}
\centering
\caption{\label{Fig2}
Extracted fit values for the quadrupole amplitude ($A_{Q}$) and the 2D widths of the NS jet-like peak ($\sigma_{\Delta\phi,NS}$, $\sigma_{\Delta\eta,NS}$) from the present analysis (stars). Left plot: Quadrupole amplitude, Middle: NS peak width on $\Delta\phi$, Right: NS peak width on $\Delta\eta$. Each plot also shows {\sc PYTHIA} predictions \cite{pythia,pythiatune} (upright triangles), dihadron results~\cite{Kettler} for $\langle p_T \rangle = 2.56$ GeV/$c$ (upside down triangles), and dihadron results for $\langle p_T \rangle = 5.7$ GeV/$c$ (dots). Horizontal bars indicate the differing centrality ranges for the present analysis and those in~\cite{Kettler}; vertical bars show the statistical errors, and cross bars show the systematic uncertainties.  }
\centering
\end{figure*}

\subsection{Model Fitting}

A quantitative representation of the $D^0 + h^{\pm}$ correlations and centrality trends is facilitated by fitting the data with a model with a minimum number of elements which are chosen to describe the visible features in the data. The $D^0 + h^{\pm}$ correlations are visually similar to previously reported, unidentified dihadron correlations; we therefore adopted the fitting model in Ref.~\cite{axialCI}. Additional and/or alternate model elements were included in the study of systematic uncertainties, discussed in the next section. We assume a NS 2D Gaussian centered at $(\Delta\eta,\Delta\phi) = (0,0)$, an AS 2D Gaussian centered at $(0,\pi)$, a $\Delta\eta$-independent quadrupole, and an overall constant offset. Both 2D Gaussians are required to be periodic on $\Delta\phi$. The model is given by
\begin{equation}
\begin{split}
& F(\Delta\eta,\Delta\phi)  =  A_0 + 2A_{\rm Q} \cos (2\Delta\phi) \\
& + A_{\rm NS} e^{-\frac{1}{2} \left[ (\Delta\eta/\sigma_{\Delta\eta,{\rm NS}})^2
+ (\Delta\phi/\sigma_{\Delta\phi,{\rm NS}})^2 \right]} \\
& + A_{\rm AS} e^{-\frac{1}{2} \left[ (\Delta\eta/\sigma_{\Delta\eta,{\rm AS}})^2
+ ((\Delta\phi - \pi)/\sigma_{\Delta\phi,{\rm AS}})^2 \right]} \\
& + {\rm periodicity},
\label{Eq9}
\end{split}
\end{equation}
where near-side Gaussian terms at $\Delta\phi = \pm 2\pi$, etc.~, and away-side Gaussians at $\Delta\phi = -\pi, \pm3\pi$, etc.,~ are not listed but are included in the model.

The eight fitting parameters $A_0$, $A_{\rm Q}$, $A_{\rm NS}$, $\sigma_{\Delta\eta,{\rm NS}}$, $\sigma_{\Delta\phi,{\rm NS}}$, $A_{\rm AS}$, $\sigma_{\Delta\eta,{\rm AS}}$, and $\sigma_{\Delta\phi,{\rm AS}}$ were, in general, allowed to freely vary to achieve the best description of the correlations based on minimum $\chi^2$. A few physically motivated restrictions were imposed, however. The quadrupole amplitude is equal to the product of the single-particle azimuthal anisotropy amplitudes $v_2^{D^0} v_2^{h^{\pm}}$, assuming factorization. Because both $v_2^{D^0} > 0$ and $v_2^{h^{\pm}} > 0$ in this collision system, the quadrupole amplitude was required to be non-negative~\cite{D0v2STAR,axialCI,cumulantv2}. For the 20-50\% and 0-20\% centrality classes, the AS Gaussian width on $\Delta\phi$ increased sufficiently that the periodic Gaussian distribution reached the dipole limit~\footnote{The cosine series representation of periodic Gaussians centered at odd-integer multiples of $\pi$ is $\sum_{k=\pm{\rm odd-integ}} \exp[-(\Delta\phi - k\pi)/2\sigma^2] = (\sigma/\sqrt{2\pi})[1+2\sum_{m=1}^{\infty} (-1)^m \exp(-m^2\sigma^2/2) \cos(m\Delta\phi)].$ For increasing $\sigma$ the series limits to the $m=1$ dipole term plus constant.}. For these two centrality classes the AS 2D Gaussian was replaced with $A_D \cos(\Delta\phi - \pi) \exp (-\Delta\eta^2/2\sigma^{2}_{\Delta\eta,{\rm AS}})$. For the cases with an undefined $\sigma_{\Delta\eta,AS}$, the fits were consistent with no AS dependence on $\Delta\eta$, and the term was therefore dropped from the fit function. Statistical fluctuations exacerbated the appearance of false, local minima in the $\chi^2$ space, resulting in false fitting solutions with unphysically narrow structures. The multi-dimensional $\chi^2$ space was mapped and necessary search limits were imposed to avoid these false solutions.

The model fits and the residuals in terms of their statistical significance, $|n\sigma_{\rm resid}|=$ $|$(data - model)/(statistical error)$|$, are shown in the middle and right-hand columns of panels in Fig.~\ref{Fig1}. The residuals are generally consistent with statistical errors except in the outermost $\Delta\eta$ bins which were omitted from the fitting procedure. The fitting model in Eq. ~\ref{Eq9} exhausts the statistically significant information in the measurements. The centrality dependences of $A_{\rm Q}$, $\sigma_{\Delta\eta,{\rm NS}}$, and $\sigma_{\Delta\phi,{\rm NS}}$, determined within the $\Delta\eta$ acceptance from $-$2 to +2 units, are shown in Fig.~\ref{Fig2}. Statistical errors and systematic uncertainties, discussed in the next section, are shown by the vertical error bars and cross-bars, respectively. The optimum fit parameters and errors are listed in Table~\ref{TableI}. The azimuthal width of the AS 2D Gaussian increases with centrality, reaching the dipole limit in the mid-central and most-central bins, providing evidence of rescattering in the medium.

\subsection{Calculating NS Associated Yield per $D^{0}$ Trigger}

The efficiency and acceptance corrected average number of associated charged particles correlated with each $D^{0}$ trigger, the {\em NS associated yield}, is approximately 

\begin{equation}
\begin{split}
& \frac{Y_{\rm NS-peak}}{N_{D^0}} \approx
\frac{dN_{\rm ch}}{2\pi d\eta} \int_{\Delta\eta \, {\rm accep}} \hspace{-0.3in} d\Delta\eta \int_{-\pi}^{+\pi} d\Delta\phi F_{\rm NS-peak}
(\Delta\eta,\Delta\phi)
\end{split}
\label{nearSideAssociateYield2}
\end{equation}
where
\begin{equation}
\begin{split}
& F_{\rm NS-peak}(\Delta\eta,\Delta\phi) = A_{\rm NS} e^{-\frac{1}{2} \left[ (\Delta\eta/\sigma_{\Delta\eta,{\rm NS}})^2
+ (\Delta\phi/\sigma_{\Delta\phi,{\rm NS}})^2 \right]}.
\end{split}
\end{equation}
The details of the derivation and calculation of the associated yield from the correlations are found in Appendix B. 
	The first term on the right hand side of Eq.~\ref{nearSideAssociateYield2} is the efficiency corrected, charged particle multiplicity in the centrality class, obtained by interpolating $dN_{ch}/d\eta$ from Table III in Ref.~\cite{axialCI} for 200 GeV Au+Au collisions. The volume of the NS correlation peak, $V_{\rm NS-peak}$ in Appendix B, is given by the integral on the right hand side where $F_{\rm NS-peak}$ in the present analysis is assumed to be the NS 2D Gaussian in Eq.~(\ref{Eq9}). Here, we are including all NS correlations, other than the quadrupole, in the NS yield per $D^0$. The $\Delta\eta$ acceptance correction factor in Appendix B, not included above, is approximately one. \

\begin{figure}[h]
\includegraphics[keepaspectratio,width=3.5in]{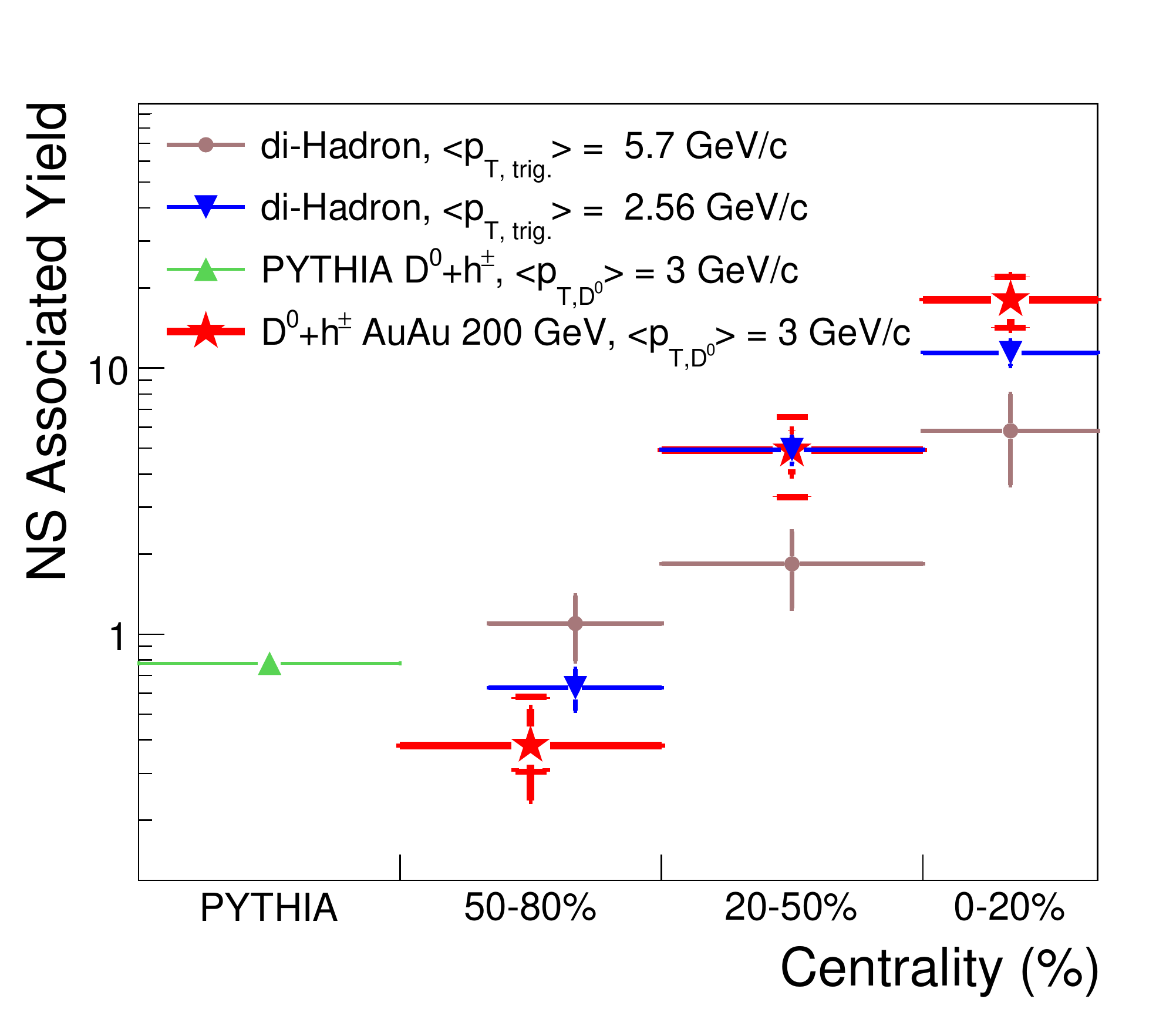}
\centering
\caption{\label{Fig4}
Correlated hadron yield per $D^0$ trigger in the near-side 2D Gaussian peak for the present data (stars), {\sc pythia}~\cite{pythia,pythiatune} predictions (upright triangle), dihadron results~\cite{Kettler} for $\langle p_T \rangle = 2.56$ GeV/$c$ (upside down triangles), and  dihadron results for $\langle p_T \rangle = 5.7$ GeV/$c$ (dots). Horizontal bars indicate the centrality ranges; vertical bars show the statistical errors, and cross bars show the systematic uncertainties.}
\centering
\end{figure}

The yield per $D^0$-trigger in the NS 2D Gaussian peak gives the average number of hadrons correlated with each $D^0$ within the acceptance. This number, shown in Fig. 5 for the assumed model description,  increases significantly with centrality as do the widths, especially the width on $\Delta\eta$, 
shown in Fig.~\ref{Fig2}. The null hypothesis, which is that the NS correlations per $D^0$-trigger are not affected by the increasing size and density of the medium, but remain constant with centrality, is strongly violated by these results (p-value $\sim 10^{-6}$).

\section{Systematic uncertainties}
\label{SecV}

We estimated systematic uncertainties in the correlation data, in the fitting model parameters, and yields per trigger. Systematic uncertainties in the 2D $D^{0}+h^{\pm}$ correlation data come from multiple sources in the analysis including: (1) variations in the $D^{0}$ topological reconstruction cuts, (2) variations in the choice of sideband widths and positions, (3) efficiency correction method, (4) $B$-meson feed-down contribution to the $D^{0}+h^{\pm}$ correlations, (5) non-primary (secondary) particle contamination, (6) uncertainty in the correction for $D^{\star \pm} \rightarrow D^{0} + \pi_{soft}$ contamination, (7) particle identification of the $D^{0}$ daughters, (8) $D^{0}$ signal and background yield estimates, (9) event-mixing multiplicity and z-vertex sub-bin widths, (10) PV position and beam+beam luminosity effects on event-wise multiplicity determination, and (11) pileup from untriggered, out-of-time collision events in the TPC. Uncertainties in the correlations were estimated for each of these sources and most were found to be either negligible or indistinguishable from statistical noise. Sources resulting in non-negligible uncertainties are discussed below along with a few others.

Systematic uncertainties (1)-(3) were estimated by varying each cut or correction individually and examining the bin-wise changes in the correlations. For error sources (2) and (3), the vast majority of the changes in the angular bins were less than 1$\sigma$ of the statistical errors, and thus not included in the systematic uncertainty. For variations in the topological cuts, error source (1) above, changes to the $D^{0}$-candidate daughter kaon and pion DCA to the PV produced larger, bin-wise changes in the correlations.  Some bins were affected by up to 4$\sigma$ in the most-central data. This non-negligible contribution was therefore included in the final systematic uncertainties. Variations in the three other topological cuts had negligible effects, and were therefore not included in the final systematic uncertainties.

Systematic uncertainties from $B$-meson feed-down to $D^{0}$-mesons were studied extensively in Ref.~\cite{D0v2STAR} where it was estimated that about 4\% of the $D^0$ sample are from feed-down. Contributions to the true, primary $D^{0}+h^{\pm}$ correlations could be as much as 4\% in the total correlation amplitude, affecting the overall normalization of the correlations. Uncertainties arising from non-primary (secondary) particle contamination were estimated in Ref.~\cite{axialCI} for unidentified charged-particle correlations. In the present analysis secondary particle contamination is suppressed for the $D^0$ candidates due to the PV resolution afforded by the HFT. The remaining contamination in the associated particle sample contributes about $\pm$1.5\% overall uncertainty in the correlation amplitudes. 

Misidentified $D^{0}$ decay daughter particles ($K \leftrightarrow \pi$) are broadly dispersed in the $M_{K\pi}$ distribution and have negligible contributions as previously reported~\cite{D0v2STAR}. Variations in the estimate of $D^{0}$ signal and background (factors $S$ and $B$) from 10\% to 20\%, as a result of changes in the background subtraction (Fig.~\ref{invariantMassDistributions}), had negligible effect on the final correlations. Requiring all tracks used in the analysis to include one hit in the IST, which resolves particles from separate beam bunch crossings~\cite{IST}, essentially eliminates pileup contamination. Event-mixing sub-bin widths were sufficiently narrow to eliminate artifacts, and the small variations in track reconstruction efficiency with PV position and luminosity negligibly affect event-wise multiplicity determination. 

Systematic uncertainty in the magnitude of the $D^{\star} \rightarrow D^0 \pi_{soft}$ contamination in the $(\Delta\eta,\Delta\phi) = (0,0)$ bin was estimated by adjusting the scale of the combinatoric background in the $M_{K\pi\pi_{\rm soft}} - M_{K\pi}$ invariant mass distribution by 20\% based on background fluctuations, causing the extracted $D^{*\pm}$ yield to vary (see Sec.~\ref{dStarSubSec}). The reduction of pairs in the $(\Delta\eta,\Delta\phi) = (0,0)$ bin from the $D^{*\pm}$ correction altered the correlations by changing the shape of the NS jet-like peak, causing a subsequent alteration of the model fit parameters. These variations are included as additional systematic uncertainties on the extracted fit parameters, and range between 3-10\%, depending on the centrality and model parameter in question.

The largest source of systematic uncertainty in the model parameters and NS peak yield is due to the choice of fitting model. The nominal fit-model was introduced in Sec. \ref{SecIV}. Systematic uncertainties were estimated by including a $\cos (3 \Delta\phi)$ (sextupole), or by replacing the $\Delta\eta$-dependent part of the NS 2D Gaussian in Eq.~(\ref{Eq9}) with either a Lorentzian function~\footnote{The Lorentzian model element is $A(\Gamma_\eta /2)^2 \exp(-\Delta\phi^2/2\sigma_{\phi}^2)/[\Delta\eta^2 + (\Gamma_\eta /2)^2].$} (leptokurtic) or a raised-cosine function~\footnote{The raised-cosine model element is $(A/2)[1 + \cos(\Delta\eta \pi/\sigma_\eta)] \exp(-\Delta\phi^2/2\sigma_{\phi}^2)$, when $|\Delta\eta| \leq \sigma_\eta$ and zero otherwise.} (platykurtic). To be included in the systematic uncertainty estimates, alternate fit models were required to have similar $\chi^2$ and residuals as the nominal fits, with one unique $\chi^{2}$-minimum corresponding to physically reasonable parameters. Finally, the nominal fitting model was applied to a rebinned version of the correlations assuming 11 $\Delta\eta$ bins and 16 $\Delta\phi$ bins, which resulted in small changes in the fit parameters. 

Each of the above positive and negative systematic uncertainties in the correlation quantities resulting from the non-negligible sources of systematic uncertainty discussed in this section were added in quadrature, where positive and negative errors were combined separately. The nominal fitting model results, statistical errors, and the combined systematic uncertainties are listed in Table~\ref{TableI}.

\begin{figure*}[t]
\includegraphics[keepaspectratio,width=6.5in]{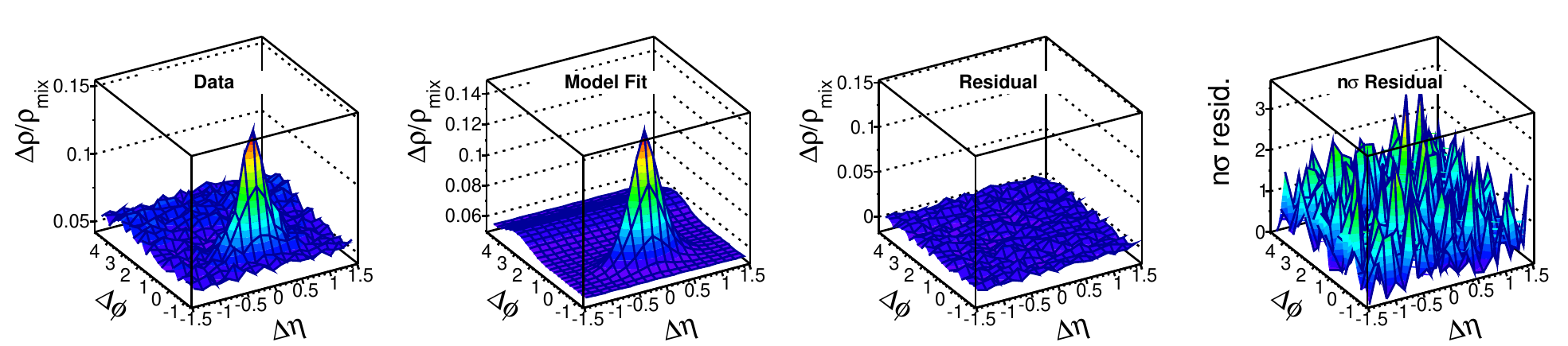}
\centering
\caption{\label{Fig5}
{\sc pythia}~\cite{pythia,pythiatune} predictions for $D^0$ + hadron 2D angular correlations in minimum-bias p+p collisions at $\sqrt{s}$ = 200~GeV in which a $D^0 \rightarrow K + \pi$ decay is produced (left panel), as well as the model fit, residuals, and $n\sigma$'s of the residuals from left to right, respectively.}
\centering
\end{figure*}

\section{Discussion}
\label{SecVI}

It is interesting to compare these results with expectations based on HF production models and on other correlation results. Angular correlations for $D^0 + h^{\pm}$ predicted by perturbative QCD with conventional fragmentation, as in {\sc pythia}~\cite{pythia,Song,pythiatune} (version 8.230)\footnote{The default tune, Monash 2013,  was used with changes to the following three parameters: 1) BeamRemnants:primordialKThard 1.8 $\rightarrow$ 1.0, 2) PhaseSpace:pTHatMin 0.0 $\rightarrow$ 1.3, and 3) TimeShower:alphaSvalue 0.1365 $\rightarrow$ 0.18, based on analyses in ~\cite{Song,pythiatune}. The decay-daughters from $K_{S}^{0}$ and $\Lambda$ were excluded from the associated track sample.}, for minimum-bias p+p collisions at $\sqrt{s}$ = 200~GeV, are shown in Fig.~\ref{Fig5} together with a model-fit (offset + NS modified 2D Gaussian~\footnote{The modified NS 2D Gaussian used to fit the {\sc pythia} correlations is $A \exp\{-[(\Delta\eta^2/2\sigma_{\Delta\eta,NS}^2)^\beta]\} \exp\{-[(\Delta\phi^2/2\sigma_{\Delta\phi,NS}^2)^\beta]\}.$} + AS Gaussian) and residuals. The correlated NS yield per trigger $D^0$ and NS peak widths are shown by the upright triangles in Figs.~\ref{Fig2} and \ref{Fig4}. The widths on $\Delta\eta$ and $\Delta\phi$ compare well with the 50-80\% centrality Au+Au results, while the NS associated yield in {\sc pythia} is approximately twice the measured yield, but is within $\sim2\sigma$ of the measurement. The NS yield per trigger and the NS $\Delta\eta$ width increase in the 20-50\% and 0-20\% centralities, relative to the perturbative QCD predictions, is similar to that reported in Ref.~\cite{axialCI} for unidentified dihadron correlations. The present centrality trends for the assumed fitting model are consistent with the onset of significant increases in the NS peak amplitude and $\Delta\eta$ width at approximately 40-50\% centrality ~\cite{axialCI}, and are consistent with the appearance of a near-side ridge in these $D^{0}$ + hadron correlations. In addition, the increase in per-trigger yield and near-side peak widths in the 20-50\% and 0-20\% centrality classes relative to the {\sc pythia} predictions occurs in the same centrality range, where a suppression for $D^{0}$ yields is observed in the same $p_{T}$ range~\cite{D0RAASTAR}. Both the present and the previous $R_{\rm AA}$ observations imply increased $D^0$ + medium interactions in more central collisions.

The reduction of the $D^0$ yield in more central collisions in the present $p_T$ range ($R_{\rm AA}$~\cite{D0RAASTAR,D0RAAALICE}) could imply that the relative yield of $D^0$ triggers emitted from the interior of the collision medium is suppressed. This resulting {\em surface bias} in the observed $D^0$ sample makes the observed increases in amplitude and width of the NS $D^0 + h^{\pm}$ correlations, in more central collisions, even more remarkable.

We also compare the $D^0 + h^{\pm}$ correlation structures with similar, unidentified charged-particle correlations for 200~GeV Au+Au collisions~\cite{Kettler} in which $p_T$ of the ``trigger'' particle is binned while the associated particle's $p_{T} \geq 0.15$~GeV/$c$. In Ref.~\cite{Kettler} the minimum-bias 200~GeV Au+Au data were divided into centrality classes 0-5\%, 5-10\%, 10-20\%, $\cdots$, 60-70\%, plus several trigger $p_T$ bins, using the same $|\eta| \leq 1$ and full $2\pi$ azimuth acceptance as in the present analysis. A similar fitting model to that used here, consisting of an offset, dipole, quadrupole and NS 2D Gaussian, was assumed in Ref.~\cite{Kettler}. The $D^0 + h^{\pm}$ and unidentified dihadron correlations are compared using a common trigger-particle $p_T$ as was assumed in Ref.~\cite{D0v2STAR} for the analysis of $D^0$ $v_2$. The light- and heavy flavor results could also be compared via a common trigger-particle velocity~\cite{Rappprivcomm} assuming that diffusion in a dispersive medium is the dominant process. The highest three trigger $p_T$ bins in Ref.~\cite{Kettler}, [2.1,3.1] GeV/$c$, [3.1,4.7] GeV/$c$ and [4.7,7.0] GeV/$c$, offer the best overlap with the $D^0$ $p_T$ range, [2,10] GeV/$c$, used here. Application of Eq.~(\ref{nearSideAssociateYield2}) to the dihadron analysis gives the approximate number of correlated charged particles per trigger-particle in the NS 2D peak as $(d\bar{N}_{\rm ch}/2\pi d\eta)V_{\rm NS-peak}$, where $\bar{N}_{\rm ch}$ is the efficiency corrected, average number of charged-particles with $p_{T} \geq 0.15$~GeV/$c$, and $V_{\rm NS-peak}$ is the volume integral of the NS 2D Gaussian within the acceptance.

Centrality fraction weighted results from Ref.~\cite{Kettler} for the 50-70\%, 20-50\% and 0-20\% centralities are shown in Figs.~\ref{Fig2} and \ref{Fig4} for the per-trigger NS 2D peak yields and for the Gaussian widths on $\Delta\phi$ and $\Delta\eta$ for trigger $p_T$ bins [2.1,3.1] GeV/$c$ ($\langle p_{T} \rangle$ = 2.56~GeV/$c$) and [4.7,7.0] GeV/$c$ ($\langle p_{T} \rangle$ = 5.7~GeV/$c$) by the upside down triangles and dot symbols, respectively. The centrality trend of the per-trigger yields for the lower-$p_T$ LF dihadron correlations follows the $D^0 + h^{\pm}$ trend fairly well; the higher $p_T$ results do not increase as rapidly with centrality. The dihadron azimuthal widths for $\langle p_{T} \rangle$ = 2.56~GeV/$c$ are similar in magnitude to the $D^0 + h^{\pm}$ widths, and follow a similar trend with centrality. The $\sigma_{\Delta\eta,{\rm NS}}$ width for $\langle p_{T} \rangle$ = 2.56~GeV/$c$ increases by a factor of 4 in the most-central bin relative to the most-peripheral bin, while the $D^0 + h^{\pm}$ results increase from peripheral to mid-centrality, but are constant within errors from mid-central to most-central.

The dihadron and $D^0 + h^{\pm}$ quadrupole amplitudes are compared in Fig.~\ref{Fig2} where the 20-50\% and 0-20\% centrality results are consistent within errors. However, the $D^0 + h^{\pm}$ quadrupole amplitude is smaller than the dihadron amplitude for peripheral collisions.

Finally, we compare the $D^0$-meson azimuthal anisotropy parameter $v_2^{D^0}$ inferred from the present analysis with a previous STAR Collaboration measurement~\cite{D0v2STAR} which used both event plane and two-particle correlation methods. The quadrupole amplitude is (see Table~\ref{TableI}) equal to $v_2^{D^0} v_2^{\rm hadron}$ assuming factorization, where $v_2^{\rm hadron}$ for 20-50\% centrality and $p_T \geq 0.15$~GeV/$c$ is approximately 0.063~\cite{axialCI}, resulting in $v_2^{D^0} \approx 0.11$ for $p_T^{D^0} \in [2,10]$~GeV/$c$. This present $v_2^{D^0}$ compares well with the previous measurement which included the 10-40\% centrality. In conventional $v_2$ analyses~\cite{D0v2STAR,D0v2ALICE} $\eta$-gaps are used to reduce the contribution of ``non-flow'' correlations, e.g. jet fragmentation. The extended near-side correlation structure on $\Delta\eta$ in the two more central bins in Fig.~\ref{Fig1} indicates that either large $\eta$-gaps or higher-order cumulant methods~\cite{cumulantv2} are required for conventional $v_2$ analyses in the more central collisions.

\section{Summary and conclusions}
\label{SecVII}

In this paper we report the first measurement of two-dimensional angular correlations between $D^0$-mesons and unidentified charged hadrons produced in relativistic heavy-ion collisions. Attention was focused on the centrality evolution of the near-side, 2D correlation peak widths and associated yields. Results for the associated hadron yield per $D^0$-meson trigger and the 2D widths of the correlated angular distribution, obtained by fitting the data with a 2D Gaussian model, were used to characterize the centrality dependence. We find that the associated hadron per $D^0$-meson yield and 2D widths increase significantly from peripheral to central collisions. With the $D^{0}$-meson serving as a proxy for a charm-quark jet, this measurement is a first attempt to understand heavy flavor jets in heavy-ion collisions at RHIC energies.

The increase in the near-side correlation yield and width coincides in both the centrality and $D^0$-meson $p_T$ ranges where the $D^0$-meson nuclear modification factor $R_{\rm AA}$ is suppressed. Both results are consistent with the expectation that the interactions between the charm quark and the medium increase with centrality. The present results complement previous studies of $D^0$-meson spectra, $R_{\rm AA}$ and $v_2$. The centrality trends and magnitudes of the NS 2D Gaussian fit parameters, qualitatively agree with a similar analysis of dihadron 2D correlations for 200~GeV Au+Au minimum-bias collisions for similar $p_T$ and centrality ranges. These results imply that the effective strength and centrality dependence of heavy flavor particle interactions with the medium are similar to that observed for light flavor particles, as seen in previous, complementary studies.

In conclusion, the near-side, non-quadrupole correlated hadrons, which are associated with $D^0$-mesons, display a large increase in per-trigger yield and 2D widths, especially the width along relative pseudorapidity, for collisions more central than about 50\%. This ridge formation phenomenon has been observed in light flavor dihadron correlations at both the RHIC and the LHC and is now observed in $D^0$-meson + hadron correlations in Au+Au collisions at 200~GeV. 

\vspace{0.1in}

{\bf Acknowledgements:}

We thank the RHIC Operations Group and RCF at BNL, the NERSC Center at LBNL, and the Open Science Grid consortium for providing resources and support.  This work was supported in part by the Office of Nuclear Physics within the U.S. DOE Office of Science, the U.S. National Science Foundation, the Ministry of Education and Science of the Russian Federation, National Natural Science Foundation of China, Chinese Academy of Science, the Ministry of Science and Technology of China and the Chinese Ministry of Education, the National Research Foundation of Korea, Czech Science Foundation and Ministry of Education, Youth and Sports of the Czech Republic, Hungarian National Research, Development and Innovation Office, New National Excellency Programme of the Hungarian Ministry of Human Capacities, Department of Atomic Energy and Department of Science and Technology of the Government of India, the National Science Centre of Poland, the Ministry  of Science, Education and Sports of the Republic of Croatia, RosAtom of Russia and German Bundesministerium fur Bildung, Wissenschaft, Forschung and Technologie (BMBF) and the Helmholtz Association.

\vspace{0.2in}

\begin{appendix}

{\bf Appendix A:} Derivation of the $D^0 + h^{\pm}$ correlation quantity. 

\vspace{0.1in}

The correlated pair distribution using all $K\pi$ parent momentum vectors from the $D^0$ signal region of the $K\pi$ invariant mass distribution, in combination with all other charged particle momentum vectors from the same event, $\Delta\rho_{\rm sig}(\Delta\eta,\Delta\phi)$, includes the following correlation sources: (1) those from true $(D^0 \rightarrow K\pi)+h^{\pm}$; (2) those from $D^0 + h^{\pm}$ where the $D^0$ meson is a decay product from the $D^{\star \pm}$ resonance and the associated charged particles exclude the soft pion daughter ($\pi_{\rm soft}$) from $D^{\star \pm}$ decay; (3) the $D^{\star \pm} \rightarrow D^0 + \pi_{\rm soft}$ pair itself, and (4) those from random combinatoric $K\pi$ plus $h^{\pm}$ pairs and from misidentified $K\pi + h^{\pm}$ pairs. The correlated pair distribution in the $D^0$ signal region can be expressed as
\bea
\Delta\rho_{\rm sig} & = & \Delta\rho_{D^0+h} + \Delta\rho_{D^0+\pi_{\rm soft}} + \Delta\rho_{{\rm BG}+h},
\label{EqA1}
\eea
where the first, second and third terms on the right-hand-side correspond to the above correlation sources (1)+(2), (3), and (4), respectively. Solving for the $D^0 + h^{\pm}$ correlations and dividing by the ideal (unobserved) $D^0 + h^{\pm}$ mixed-event distribution gives
\bea
\frac{ \Delta\rho_{D^0+h}}{\alpha \rho_{{\rm ME},D^0+h}} & = &
\frac{\Delta\rho_{\rm sig} - \Delta\rho_{D^0+\pi_{\rm soft}} - \Delta\rho_{{\rm BG}+h}}
     {\alpha \rho_{{\rm ME},D^0+h}}.
\label{EqA2}
\eea
The first two terms on the RHS can be rewritten as
\bea
\frac{\Delta\rho_{\rm sig} - \Delta\rho_{D^0+\pi_{\rm soft}}}{\alpha \rho_{{\rm ME},D^0+h}} &=&
\nonumber \\
  &  & \hspace{-0.5in} = \frac{\alpha \rho_{\rm ME,sig}}{\alpha \rho_{{\rm ME},D^0+h}}
\frac{\Delta\rho_{\rm sig} - \Delta\rho_{D^0+\pi_{\rm soft}}}{\alpha \rho_{\rm ME,sig}} 
\nonumber \\
 &  &  \hspace{-0.5in} =  
\frac{S+B}{S} 
\frac{\Delta\rho_{\rm sig} - \Delta\rho_{D^0+\pi_{\rm soft}}}{\alpha \rho_{\rm ME,sig}}
\label{EqA3}
\eea
and the third term becomes
\bea
\frac{\Delta\rho_{{\rm BG}+h}}{\alpha \rho_{{\rm ME},D^0+h}} & = & 
\frac{\alpha \rho_{\rm ME,BG}}{\alpha \rho_{{\rm ME},D^0+h}}
\frac{\Delta\rho_{{\rm BG}+h}}{\alpha \rho_{\rm ME,BG}}
\nonumber \\
 & = &
\frac{B}{S}
\frac{\Delta\rho_{{\rm BG}+h}}{\alpha \rho_{\rm ME,BG}}.
\label{EqA4}
\eea
Substituting these results into Eq.~(\ref{EqA2}) and using the average of the left and right side-bands to estimate the background correlations, gives the final expression in Eq.~(\ref{Eq1}) of the main text:
\bea
 \frac{\Delta\rho_{D^0+h}}{\alpha \rho_{{\rm ME},D^0+h}}  & = &  
\frac{S+B}{S} \frac{\Delta\rho_{\rm sig} - \Delta\rho_{D^0 + \pi_{\rm soft}}}
                   {\alpha \rho_{\rm ME,sig}}
\nonumber \\
  & -  &  \frac{B}{S} \frac{\Delta\rho_{\rm SB}}{\alpha_{\rm SB} \rho_{\rm ME,SB}}.
\eea

\end{appendix}

\vspace{0.2in}

\begin{appendix}

{\bf Appendix B:} Calculation of the near-side correlated yield per $D^0$ trigger.

\vspace{0.1in}

The correlated pair yield per $D^0$ trigger in the NS 2D peaked correlation structure, $Y_{\rm NS-peak}/N_{D^0}$, is estimated by summing that portion of the correlation fitting model in Eq.~(\ref{Eq9}) over the $(\Delta\eta,\Delta\phi)$ acceptance, including efficiency and acceptance corrections, and dividing by the efficiency corrected number of $D^0$ mesons, $N_{D^0}$, used in the analysis. This estimate is given by~\cite{ALICEyields}
\bea
Y_{\rm NS-peak}/N_{D^0} & = & \frac{1}{N_{D^0}}
\nonumber \\
 & & \hspace{-0.8in} \times \sum_{\Delta\eta,\Delta\phi}
\left[ \frac{\Delta n_{D^0 + h}} {\frac{\alpha \rho_{{\rm ME},D^0+h}}{\alpha \rho^{\rm max}_{{\rm ME},D^0+h}}} \right]_{\rm NS-peak},
\label{EqB1}
\eea
where $\Delta n_{D^0 + h} = \delta_{\Delta\eta} \delta_{\Delta\phi} \Delta\rho_{D^0+h}$ in each $(\Delta\eta,\Delta\phi)$ bin, and $\delta_{\Delta\eta},\delta_{\Delta\phi}$ are the bin widths on $\Delta\eta$ and $\Delta\phi$. Also in Eq.~(\ref{EqB1}), $\alpha \rho^{\rm max}_{{\rm ME},D^0+h}$ is the maximum value of the normalized, mixed-event pair distribution, evaluated by averaging over the $\Delta\phi$ bins for $\Delta\eta = 0$. The ratio in the denominator represents the detector acceptance distribution normalized to 1.0 at the maximum. Rearranging Eq.~(\ref{EqB1}) gives
\bea
Y_{\rm NS-peak}/N_{D^0} & = & \frac{\alpha \rho^{\rm max}_{{\rm ME},D^0+h}}{N_{D^0}}
\nonumber \\
 & & \hspace{-.5in} \times \sum_{\Delta\eta,\Delta\phi} \delta_{\Delta\eta} \delta_{\Delta\phi}
\left[ \frac{\Delta \rho_{D^0 + h}} {\alpha \rho_{{\rm ME},D^0+h}} \right]_{\rm NS-peak}
\nonumber \\
 & & \hspace{-.5in} = \frac{\alpha \rho^{\rm max}_{{\rm ME},D^0+h}}{N_{D^0}} V_{\rm NS-peak}
\label{EqB2}
\eea
where the summation in the second line of Eq.~(\ref{EqB2}) is defined as $V_{\rm NS-peak}$, the volume of the NS peak correlation structure.

The ratio on the RHS of the third line of Eq.~(\ref{EqB2}) can be estimated from the measured numbers of $D^0$ and $D^0 + h^{\pm}$ ME pairs, provided both numerator and denominator are corrected for inefficiencies. A simpler form is given in the following in which the required efficiency corrected quantities are more readily obtained.

The maximum value of the efficiency corrected, normalized mixed-event density equals the fraction of the total number of $D^0 + h^{\pm}$ pairs in a $\Delta\eta = 0$, $\Delta\phi$ bin per bin area. This is given by
\bea
\alpha \rho^{\rm max}_{{\rm ME},D^0+h} & = & \frac{\varepsilon \bar{n}_{D^0} \bar{n}_h} {\delta_{\Delta\eta} \delta_{\Delta\phi}} \frac{2[1 - 1/(2N_{\Delta\eta})]} {N_{\Delta\eta} N_{\Delta\phi}}
\label{EqB3}
\eea
where $\varepsilon$ is the number of events in the centrality class, $\bar{n}_{D^0}$ and $\bar{n}_h$ are the efficiency corrected, event-averaged number of $D^0$ mesons and associated $h^{\pm}$ particles in the acceptance, $N_{\Delta\eta}$ and $N_{\Delta\phi}$ are the numbers of $\Delta\eta$ and $\Delta\phi$ bins, where $N_{\Delta\eta}$ is odd and $N_{\Delta\phi}$ is a multiple of four. The second ratio on the RHS of Eq.~(\ref{EqB3}) is the fraction of $D^0 + h^{\pm}$ pairs in an average $\Delta\eta = 0$, $\Delta\phi$ bin. The efficiency corrected number of $D^0$ mesons is $N_{D^0} = \varepsilon\bar{n}_{D^0}$. The ratio in Eq.~(\ref{EqB2}) simplifies to
\bea
\frac{\alpha \rho^{\rm max}_{{\rm ME},D^0+h}}{N_{D^0}} & = & \frac{2 \bar{n}_h  [1 - 1/(2N_{\Delta\eta})]}
{N_{\Delta\eta} N_{\Delta\phi} \delta_{\Delta\eta} \delta_{\Delta\phi}}
\nonumber \\
 & = & \frac{\bar{n}_h}{2\pi \Omega_{\eta}} \left( 1 - \frac{1}{2 N_{\Delta\eta}} \right)
\nonumber \\
 & \approx & \frac{dN_{\rm ch}}{2\pi d\eta} \left( 1 - \frac{1}{2 N_{\Delta\eta}} \right)
\label{EqB4}
\eea
where $N_{\Delta\eta} N_{\Delta\phi} \delta_{\Delta\eta} \delta_{\Delta\phi} = 4\pi \Omega_\eta$, and $\Omega_\eta$ is the single particle pseudorapidity acceptance which equals 2 units for the STAR TPC~\cite{STARNIM}. In the last step we assumed that the number of $K,\pi$ daughters is much less than the event multiplicity, such that $\bar{n}_h$ is well approximated by event multiplicity $N_{\rm ch}$. The final NS-peak correlated yield per $D^0$ trigger is given by
\bea
Y_{\rm NS-peak}/N_{D^0} & = & \frac{dN_{\rm ch}}{2\pi d\eta} \left( 1 - \frac{1}{2 N_{\Delta\eta}} \right)
V_{\rm NS-peak}
\label{EqB5}
\eea
where $dN_{\rm ch}/2\pi d\eta$ is efficiency corrected~\cite{axialCI}.

\end{appendix}


\end{document}